\documentclass[11pt]{article}
\usepackage{graphicx}
\usepackage{multirow}
\usepackage[authoryear]{natbib}
\usepackage{amsmath, amssymb}
\usepackage[utf8]{inputenc}
\usepackage{amsthm}
\usepackage{subfigure}
\usepackage{color}
\usepackage[T3,OT1]{fontenc}
\usepackage[colorlinks,bookmarksopen,bookmarksnumbered,citecolor=blue,urlcolor=green,hypertexnames=false]{hyperref}
\DeclareSymbolFont{tipa}{T3}{cmr}{m}{n}
\DeclareMathAccent{\invbreve}{\mathalpha}{tipa}{16}
\usepackage{setspace}
\usepackage{geometry}
\begin{document}
\title{Multidimensional Adaptive Penalised Splines with Application to Neurons' Activity Studies}%
\author{Mar\'ia Xos\'e Rodr\'iguez - \'Alvarez$^{1,2}$, Mar\'ia Durb\'an$^{3}$, Paul H. C. Eilers$^{4}$, \\Dae-Jin Lee$^{1}$, Francisco Gonzalez $^{5}$\\\\       
				\small{$^{1}$ BCAM - Basque Center for Applied Mathematics, Bilbao, Spain}\\
				\href{mailto:mxrodriguez@bcamath.org}{mxrodriguez@bcamath.org}\\
				\small{$^{2}$ IKERBASQUE, Basque Foundation for Science, Bilbao, Spain}\\
        \small{$^{3}$ Department of Statistics, Universidad Carlos III de Madrid, Legan\'es, Spain}\\
        \small{$^{4}$ Erasmus University Medical Centre, Rotterdam, the Netherlands}\\                
		\small{$^{5}$ Department of Surgery and CiMUS, University of Santiago de Compostela, Spain}}
\maketitle
\date{}
\sloppy
\begin{abstract}
P-spline models have achieved great popularity both in statistical and in applied research. A possible drawback of P-spline is that they assume a smooth transition of the covariate effect across its whole domain. In some practical applications, however, it is desirable and needed to adapt smoothness locally to the data, and adaptive P-splines have been suggested. Yet, the extra flexibility afforded by adaptive P-spline models is obtained at the cost of a high computational burden, especially in a multidimensional setting. Furthermore, to the best of our knowledge, the literature lacks proposals for adaptive P-splines in more than two dimensions. Motivated by the need for analysing data derived from experiments conducted to study neurons' activity in the visual cortex, this work presents a novel locally adaptive anisotropic P-spline model in two (e.g., space) and three (space and time) dimensions. Estimation is based on the recently proposed SOP (Separation of Overlapping Precision matrices) method, which provides the speed we look for. The practical performance of the proposal is evaluated through simulations, and comparisons with alternative methods are reported. In addition to the spatio-temporal analysis of the data that motivated this work, we also discuss an application in two dimensions on the absenteeism of workers.
\end{abstract}

\section{Introduction}
The estimation of curves or surfaces using penalised splines \citep{OSullivan1986, Eilers1996} has become one of the most popular methods in non-parametric regression. Penalised splines combine low-rank basis with a penalty term. This term controls the smoothness of the estimated function and its influence is determined by a smoothing parameter. In the \emph{classic} penalised spline approach this parameter is global in the sense  that provides a constant amount of smoothing. This might be a serious drawback in situations where the features of the functions present strong heterogeneity. This happens, for example, in signal regression when the function changes rapidly in some areas and is smooth in others, or in spatial data that present non-stationarity. In the case of univariate smoothing, a wide range of solutions have been proposed. \cite{Ruppert00} based their approach on a varying penalty and a multivariate cross-validation approach for the selection of the smoothing parameters. \cite{Krivobokova08} propose a model with varying smoothing parameter modeled as another penalised spline formulated as a hierarchical mixed model, and use a Laplace approximation of the marginal likelihood for parameter estimation. More recently, \cite{Yue14} introduced a class of adaptive smoothing spline models that is derived by solving certain stochastic differential equations with finite element methods. The task  becomes more complicated when trying to achieve adaptivity in two dimensions. As far as we are aware, all attempts to  two-dimensional adaptive smoothing have been proposed in the context of isotropic smoothing, i.e., the same amount of smoothing is used in both dimensions. For example, \cite{Lang04} use locally adaptive smoothing parameters which are incorporated using a smoothness prior with spatially adaptive variances and  \cite{Yue10} improves this work by proposing a prior with a spatially adaptive variance component and taking a further Gaussian Markov Random Field prior for this variance function. A different approach is taken by  \cite{Jang11} which reparametrised the smoothing parameter as a smooth step function that can be extended to higher dimensions, \cite{Krivobokova08} achieve spatial adaptivity by imposing a functional structure on the smoothing parameters in ordinary penalised splines, and \cite{Reiss17} uses adaptive smoothing and a non-isotropic penalty, but the adaptivity is only in one direction, allowing only one smoothing parameter to vary along the other dimension. However, all these approaches, are computationally very demanding, unstable, or do not include general smoothing structures (such as anisotropic smoothing).

Motivated by the need for analysing neurons' activity in the visual cortex, this work presents a general framework for anisotropic multidimensional adaptive smoothing in the context of P-spline models \citep{Eilers1996}, where B-spline basis are combined with a discrete penalty on the coefficients. Our proposal relies on the construction of locally adaptive penalties through assuming a different smoothing parameter for each coefficient difference in the penalty. Its use is not restricted to Gaussian responses (as it is in the case of most of the existing approaches) since it can be easily extended to responses within the exponential family of distributions. One of the possible drawbacks of adaptive smoothing is the computational cost of having to estimate (or select by cross-validation methods or information criteria) multiple smoothing parameters. We solve this issue by using the equivalence between P-splines and generalised linear mixed models \citep{Currie2002} and the recently developed SOP (Separation of Overlapping Penalties) method \citep{MXRA19}. Furthermore, we will show that, even when adaptive smoothing is not needed, the method is still efficient since it can make better use of the degrees of freedom by setting to zero the unnecessary coefficients. 

The rest of the paper is organised as follows: in Section \ref{motivation} we introduce the neuron' activity study that motivated this work. Section \ref{multi_penalty} presents in detail the construction of locally adaptive penalties in two dimensions and its extension to the three-dimensional case. Details of the estimation and computational aspects are given in Section \ref{Estimation}. The empirical performance of the approach is evaluated in a simulation study in Section \ref{simulation}. In section \ref{Application}, the application to the neurons' activity study, and to another data set, is shown, and we conclude with a discussion. Extended simulations are available as Supplementary Material.
\section{Motivating Example: Neurons' Activity Study}\label{motivation}
\begin{figure}[h!]
 \begin{center}
 \includegraphics[width=15cm]{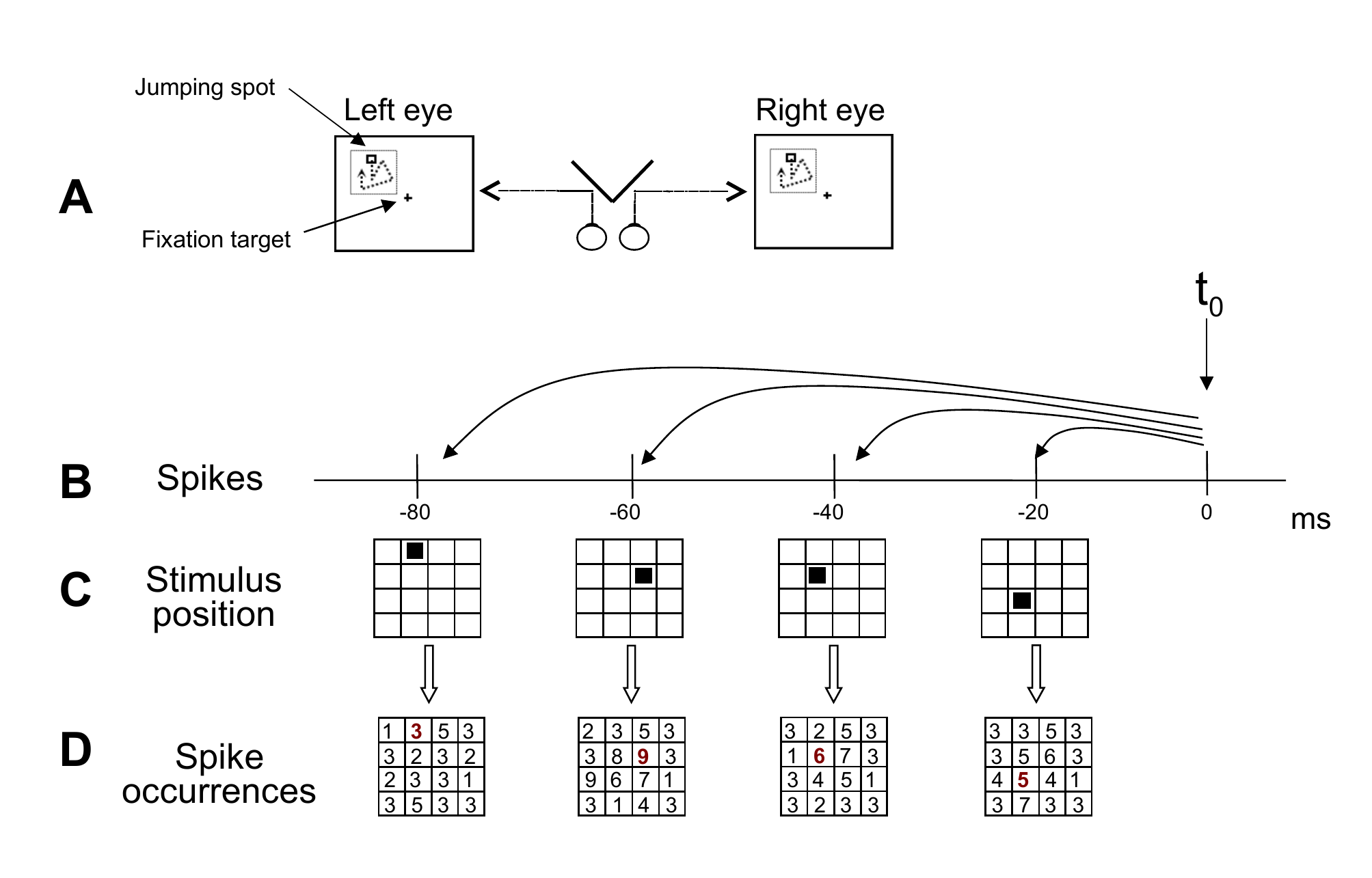}
	\end{center}
	\caption{Reverse cross-correlation technique. The animal is viewing two monitors (A) with a fixation target. Within a square area over the cell visual field a bright or dark spot is flashed in different positions in a pseudo-random manner. Cell spikes are recorded while the stimulus is delivered (B). When a spike is produced ($t_0$), the stimulus position at several pre-spike times ($-20, -40, \ldots , -320$ ms) is read (C) and the corresponding position is increased by one (D).} %The result is a set of $16$ numerical matrix containing the number of stimulus occurrences at each position and pre-spike time (D).
	\label{cross_correlation}
\end{figure}

\begin{figure}[h!]
 \begin{center}
 \includegraphics[width=12cm]{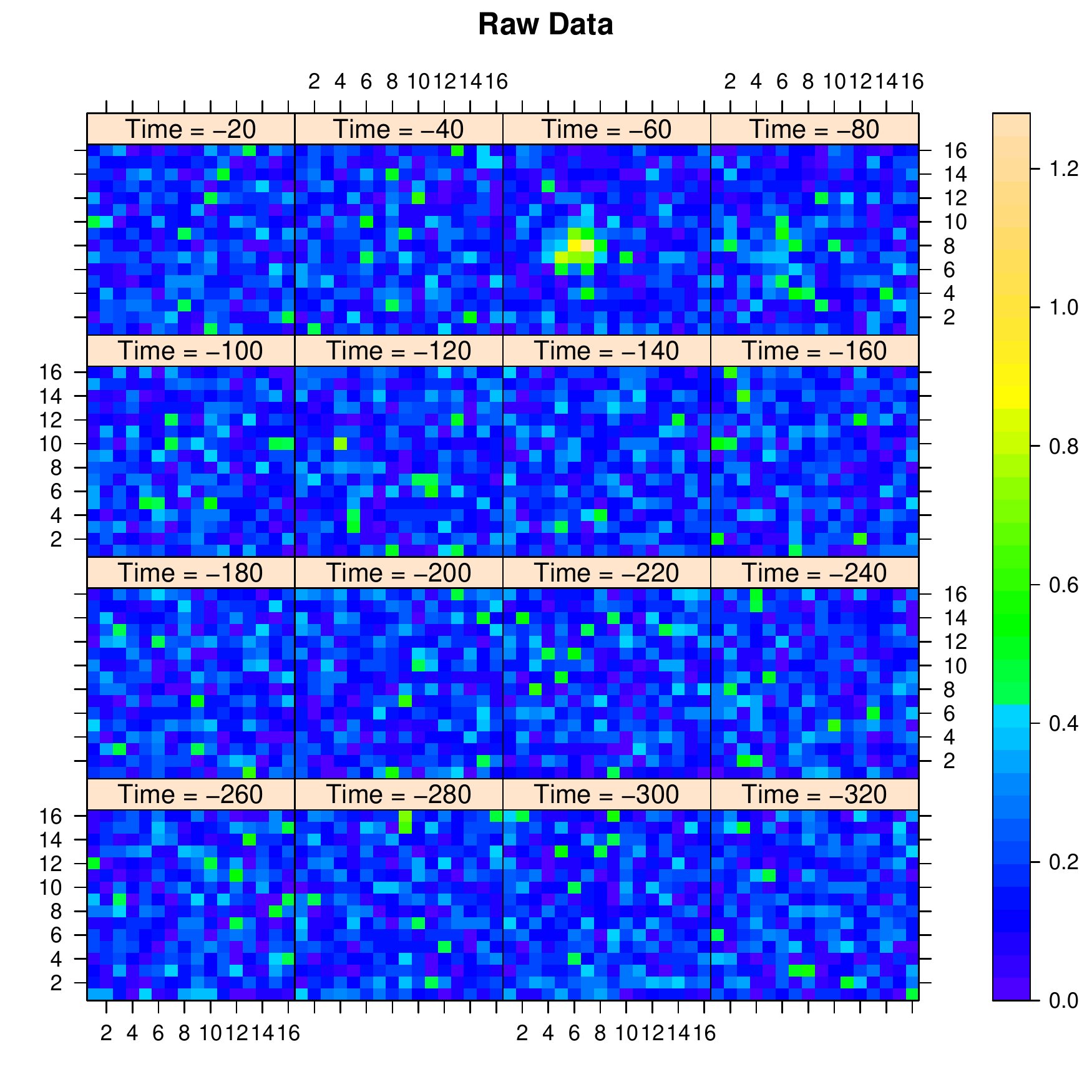}
	\end{center}
	\caption{For the visual receptive field study: Level plot of the observed ON-RFmaps (firing rates) for the right eye of cell FAU3.} %The result is a set of $16$ numerical matrix containing the number of stimulus occurrences at each position and pre-spike time (D).
	\label{vrf_rawdata}
\end{figure}

The work described in this paper was motivated by the research and electrophysiology studies conducted by Francisco Gonzalez, professor of Ophthalmology at the University of Santiago de Compostela (Galicia, Spain) and co-author of the work. The experiment we deal with here was conducted to study neurons' activity in the visual cortex, in particular, what is called visual receptive fields (RFs). RFs are small areas of the visual field that a particular visual neuron ``sees'' (i.e., areas that elicit neuronal responses). They can be mapped using different techniques, such a reverse cross-correlation. In particular, these techniques allow studying how visual neurons process signals (sensory stimuli) from different positions in their visual field. Since as a result of a sensory stimulus, a neuron can produce sudden changes in its membrane potential known as ``spikes'', from the neuron responses (spikes) it is possible to infer the spatio-temporal properties of the RFs (i.e., when and where a sensory stimulus produces a response). A detailed explanation of the electrophysiological experiment and the reverse cross-correlation technique discussed here can be found in \cite{Rodriguez2012} (and references therein). That paper was also the starting point of this work. Schematically, the experiment is as follows (Figure \ref{cross_correlation}). The subject is viewing two monitors, one for each eye. In each monitor, there is a square area of dimension $16 \times 16$ (i.e. $256$ spatial/grid locations). In a pseudo-random way, stimuli are delivered at these spatial locations (Figure \ref{cross_correlation}A). The experiment is conducted under two different experimental conditions. Namely, the stimulus can correspond to the flash of a bright (``ON'') or dark (``OFF'') spot. While the stimuli are being delivered, the activity of the neuron is being recorded (Figure \ref{cross_correlation}B). When a spike occurs, say at $t_0$, the location of the stimulus at different pre-spike times ($-20, -40, \ldots , -320$ milliseconds (ms); for this experiment it is not expected that a stimulus would produce a response (spike) after 320 ms) is recovered (Figure \ref{cross_correlation}C). Unfortunately, the way the experiment is performed does not allow to know which of these stimuli (and thus locations) is responsible for the spike (neuron's response), and all are considered as potentially responsible. As such, the number of spike occurrences attributed to the location of the stimulus at the different pre-spike times is increased by one (Figure \ref{cross_correlation}D). However, since, during the experiment, stimuli are randomly delivered/presented many times, this allows determining where and when a stimulus produces a neuron's response. Summarising, for each neuron (this study contains data for $17$ different neurons), eye (left or right) and experimental condition (``ON'' or ``OFF''), the reverse cross-correlation technique provides a dataset consisting of a series of $16$ matrices (one for each pre-spike time) of dimension $16\times 16$ ($256$ grid positions that represent the square area). Each cell of each of the $16$ matrices contains the number of spike occurrences attributed to this location at the corresponding pre-spike time. Besides, there is an extra matrix of dimension $16\times 16$ with the number of stimulus presentations on each particular location of the square area. The graphical representation of each of the $16$ matrices (normalised, i.e., each cell is divided by the number of stimulus presentations in this location) is called receptive field map (RFmap) and can be regarded as a representation of the firing rate of the neuron. Figure \ref{vrf_rawdata} depicts the evolution over time of the RFmap for a particular neuron (denoted by FAU3), eye (right) and experimental condition (``ON''). Despite the noisy data, it can be observed that for most pre-spike times the firing rate is uniform, but there is a clear increase in the firing rate around $-60$ ms for a central area of the visual field. In other words, the results suggest that only stimuli in this area produce a response of the neuron and that the response occurs around $60$ ms after the stimulus is presented. This area corresponds to the RF of the neuron. The work to be presented in this paper aims to use P-spline models to provide smoothed (de-noised) versions of RFmaps such as those shown in Figure \ref{vrf_rawdata}. Yet note that the RFmaps show that the transition from outside the RF into the RF is very sharp and that they are structured for a short period (between $-40$ and $-80$ ms). This suggests the need for multidimensional adaptive smoothing, which was also pointed out in the discussion of \cite{Rodriguez2012}'s paper. Our approach is described in the next Section, and results for this study are discussed in Section \ref{app_vrc}. Although this will not be covered here, we note that de-noised RFmaps will facilitate the study of several characteristics of RFs, such as their size or centre, as well as comparisons among neurons and/or experimental conditions.

\section{Multidimensional Adaptive Penalty}\label{multi_penalty}
This section presents our proposal for the construction of locally adaptive penalties in more than one dimension. Our approach builds upon the same principles as the adaptive penalty in one dimension \citep[see, e.g.,][and references therein]{MXRA19}. In consequence, to make the presentation of the new results more readable, we first briefly focus on the one-dimensional case and then we move onto the multidimensional setting. For clarity, the section describes in full detail the rationale for the adaptive penalty in two dimensions and ends with the generalisation to the three-dimensional case. %In all cases, we first describe the ``standard'' P-spline model and associated penalty, and then  For clarity, here we describe in full detail the rationale of the adaptive penalty in two dimensions, and end the section with the generalisation to the three dimensional case. We note that the locally adaptive penalty in one dimension has been dealt with in detail elsewhere \citep[see, e.g.,][]{MXRA19}. 
\subsection{Adaptive Penalty in One Dimension}\label{MX:adapt_pen_1D}
\begin{figure}
 \begin{center}
 \subfigure[Standard penalty]{\includegraphics[width=12cm]{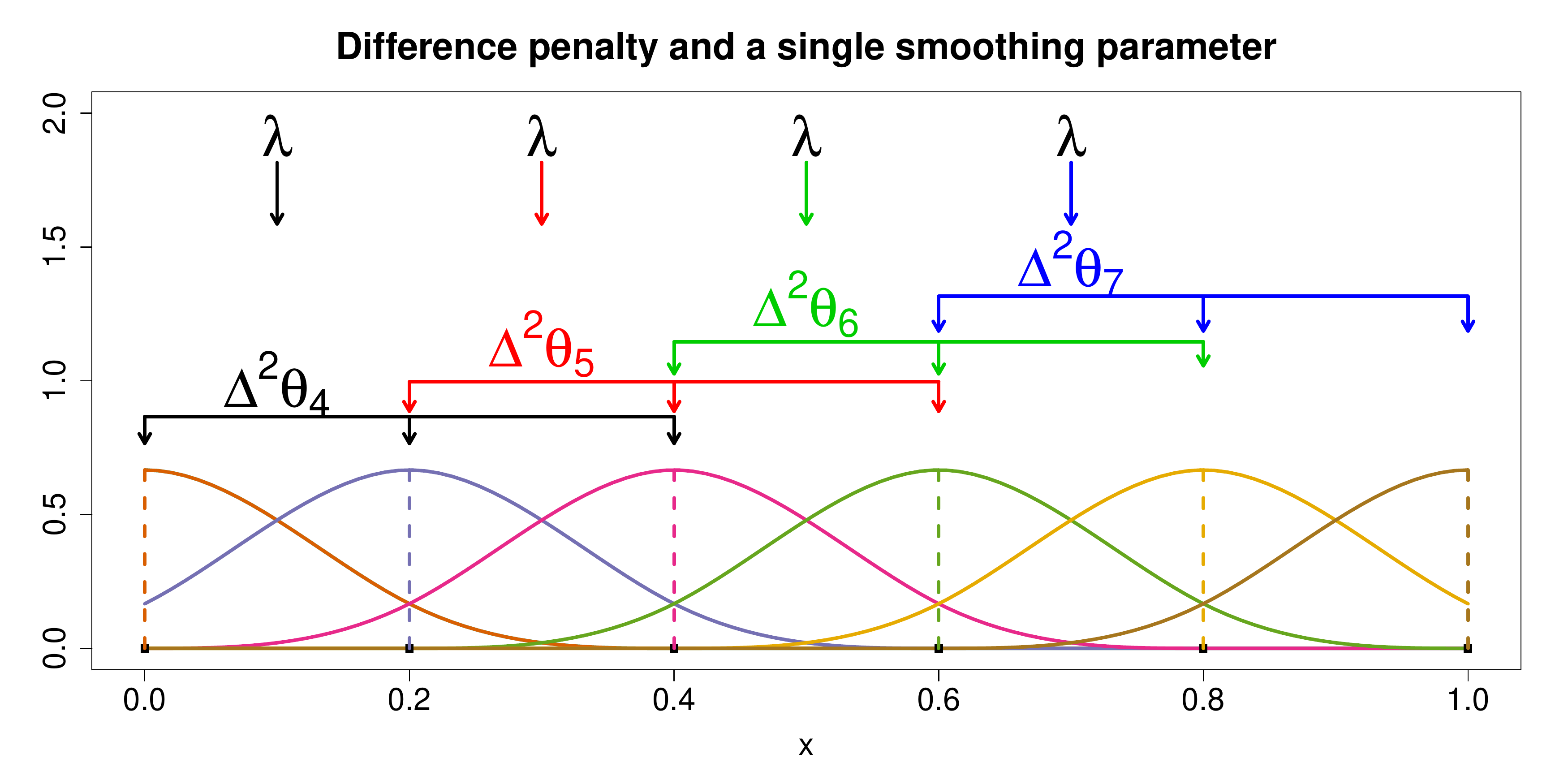}\label{diff_penalty_1d}}
 \subfigure[Locally adaptive penalty]{\includegraphics[width=12cm]{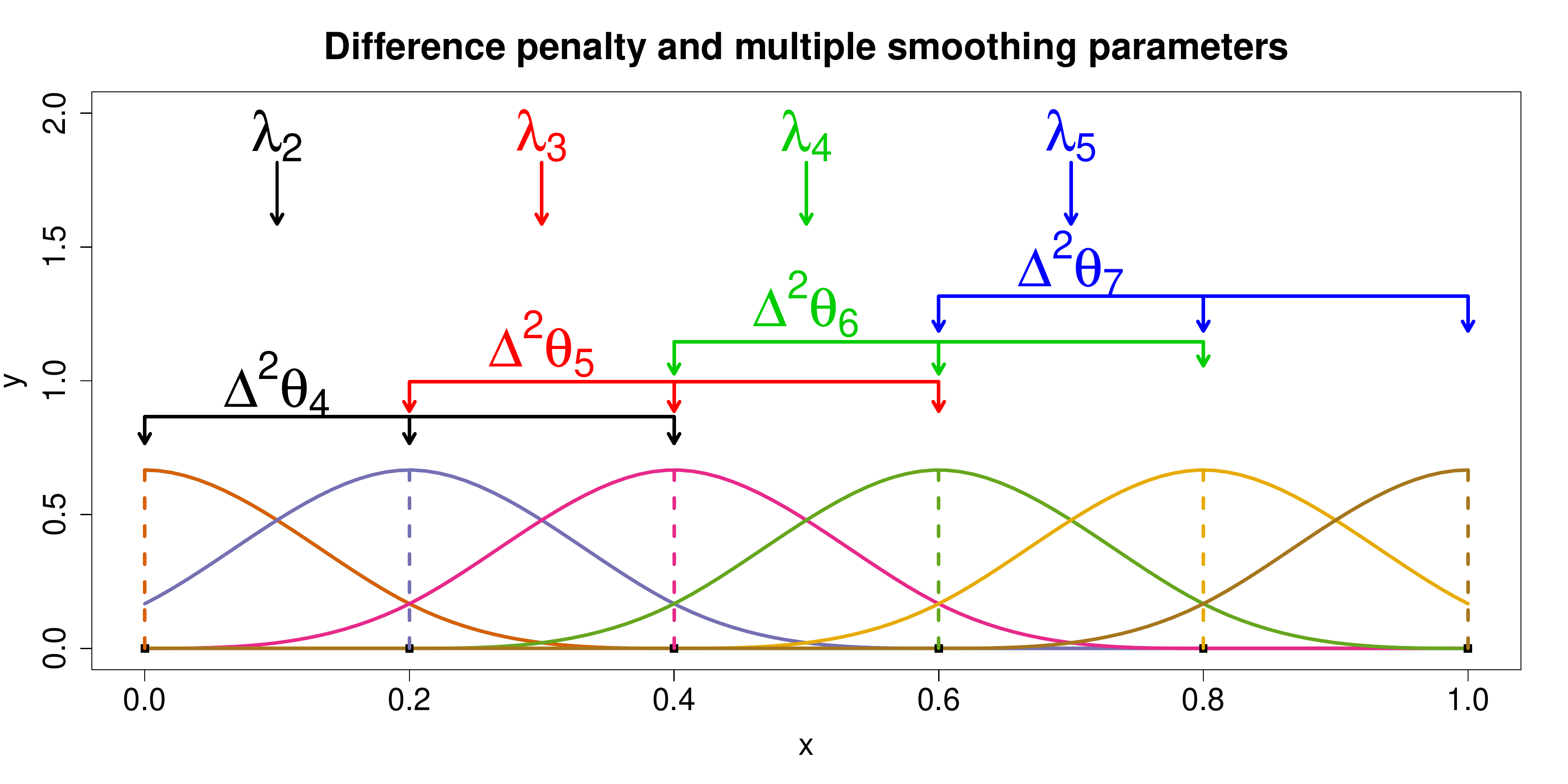}\label{diff_penalty_1d_add}}
	\end{center}
	\caption{Graphical representation of differences of order 2 on adjacent coefficients of cubic B-splines basis functions. Top row: Standard penalty; one smoothing parameter for all coefficient differences. Bottom row: Locally adaptive penalty; a different smoothing parameter for each coefficient difference.}
	\label{diff_penalty}
\end{figure}

Let $\boldsymbol{y} = \left(y_1,\ldots,y_n\right)^{\top}$ be a vector of $n$ observations, and consider the (simple) generalised model
\begin{equation*}
g\left(\mu_i\right) = f\left(x_i\right), \;\;\;\;\; i = 1,\ldots n,
\label{MX:PsplineM}
\end{equation*}
where $\mu_i = \mathbb{E}\left(y_i \mid x_i\right)$, $\mathbb{V}\mbox{ar}\left(y_i \mid x_i\right) = \phi\nu\left(\mu_i\right)$, $g(\cdot)$ is the link function, and $f(\cdot)$ is a smooth and unknown function. Here, $\nu\left(\cdot\right)$ is a specified variance function, and $\phi$ is the dispersion parameter that may be known or unknown. In P-splines \citep{Eilers1996}, the function $f(x)$ is modelled as a linear combination of B-splines basis functions, i.e., $f(x) = \sum_{j=1}^{d}\theta_{j}B_{j}\left(x\right)$, and smoothness is ensured by penalising the differences of order $q$ of coefficients associated with adjacent B-spline basis functions, i.e., the penalty takes the following form
\begin{equation}
\lambda\sum_{k = q + 1}^{d}\left(\Delta^q\theta_k\right)^2 = \lambda \boldsymbol{\theta}^{\top}\boldsymbol{D}_q^{\top}\boldsymbol{D}_q\boldsymbol{\theta},
\label{MX:1DPenalty}
\end{equation}
where $\Delta^q$ forms differences of order $q$, $\boldsymbol{\theta} = \left(\theta_1, \theta_2, \ldots, \theta_d\right)^{\top}$ and $\boldsymbol{D}_q$ is the matrix representation of $\Delta^q$. Finally, $\lambda$ is the smoothing parameter that controls the trade off between fidelity to the data (when $\lambda$ is small) and smoothness of the function estimate (when $\lambda$ is large). Note that \eqref{MX:1DPenalty} penalises all coefficient differences, $\Delta^q\theta_k$ ($k = q+1, \ldots, d$), by the same smoothing parameter $\lambda$ (see Figure \ref{diff_penalty_1d}). Implicit in equation.~(\ref{MX:1DPenalty}) is thus the assumption that the same amount of smoothing is needed across the whole domain of the covariate. The locally adaptive penalty relaxes this assumption by assuming a different smoothing parameter for each coefficient difference   
\begin{equation}
\sum_{k = q + 1}^{d}\lambda_{k-q}\left(\Delta^q\theta_k\right)^2 = \boldsymbol{\theta}^{\top}\boldsymbol{D}_q^{\top}\mbox{diag}(\boldsymbol{\lambda})\boldsymbol{D}_q\boldsymbol{\theta},
\label{MX:AdaptPenalty}
\end{equation}
where $\boldsymbol{\lambda} = \left(\lambda_1,\ldots,\lambda_{d-q}\right)^{\top}$ is a vector of smoothing parameters. That is, the adaptive penalty defined in \eqref{MX:AdaptPenalty} allows the amount of smoothing (driven by the smoothing parameters $\boldsymbol{\lambda}$) to vary locally depending on the covariate values. This is graphically illustrated in Figure \ref{diff_penalty_1d_add}. To reduce the complexity of the adaptive penalty in \eqref{MX:AdaptPenalty} (there are as many smoothing parameters as coefficient differences, i.e., $d - q$), the vector of smoothing parameters $\boldsymbol{\lambda}$ is further replaced by a smooth version of it $\boldsymbol{\xi} = \left(\xi_1,\ldots,\xi_{p}\right)^{\top}$ (with $p < (d-q)$ so as to ensure that the number of smoothing parameters is reduced) using a B-spline basis expansion, i.e.,
\begin{equation}
\lambda_k = \sum_{l = 1}^{p}\xi_l\psi_l(k).
\label{MX:lambda_be}
\end{equation}
Plugging-in the right-hand side of previous equation into \eqref{MX:AdaptPenalty}, the locally adaptive penalty is expressed as
\begin{equation}
\boldsymbol{\theta}^{\top}\left(\sum_{l=1}^{p}\xi_{l}\boldsymbol{D}_q^{\top} \mbox{diag}\left(\boldsymbol{\psi}_{l}\right)\boldsymbol{D}_q\right)\boldsymbol{\theta},
\label{MX:AdaptPenalty_smooth}
\end{equation}
where $\boldsymbol{\psi}_{l} = \left(\psi_l(1), \psi_l(2), \ldots, \psi_l(d-q)\right)^{\top}$. Note that, in matrix form, \eqref{MX:lambda_be} is written as
\begin{equation}
\boldsymbol{\lambda} = \boldsymbol{\Psi}\boldsymbol{\xi},
\label{MX:Adapt_smooth}
\end{equation}
where
\[
\boldsymbol{\Psi} = \left(
\begin{array}{ccc}
\psi_{1}(1) & \ldots  & \psi_{p}(1)\\
\vdots & \ddots & \vdots\\
\psi_{1}(d-q) & \ldots  & \psi_{p}(d-q)
\end{array}
\right),
\]
is the B-spline design  matrix of dimension $\left(d-q\right) \times p$, and thus $\boldsymbol{\psi}_{l}$ in \eqref{MX:AdaptPenalty_smooth} is the column $l$ of $\boldsymbol{\Psi}$.%, the B-spline design matrix.
\subsection{Adaptive Penalty in Two Dimensions}\label{sec_2D}
\begin{figure}[!htp]
 \begin{center}
 %\subfigure[Landscape of nine cubic B-spline tensor products, a portion of a full basis]{
 \includegraphics[width=7.2cm]{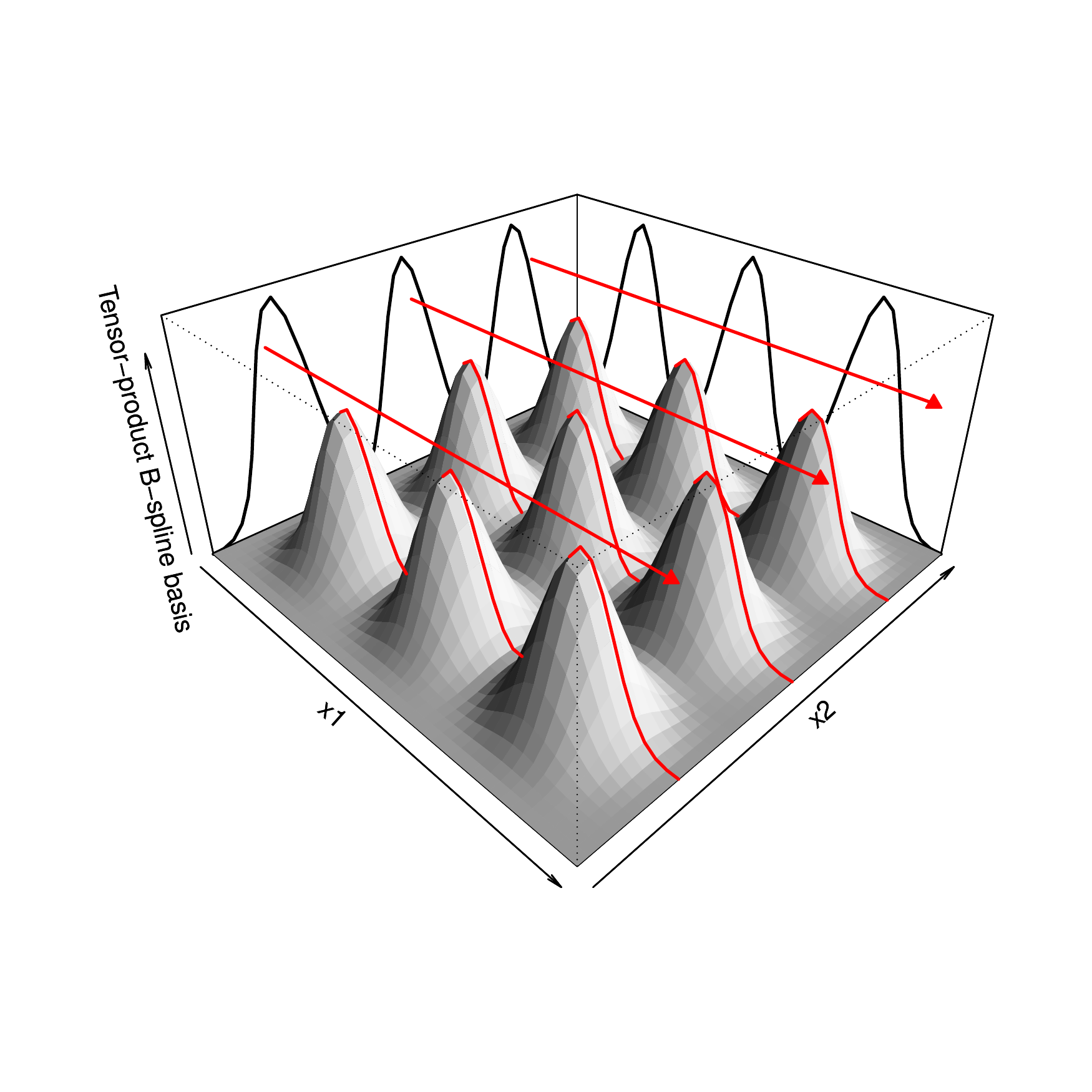}
 \includegraphics[width=7.2cm]{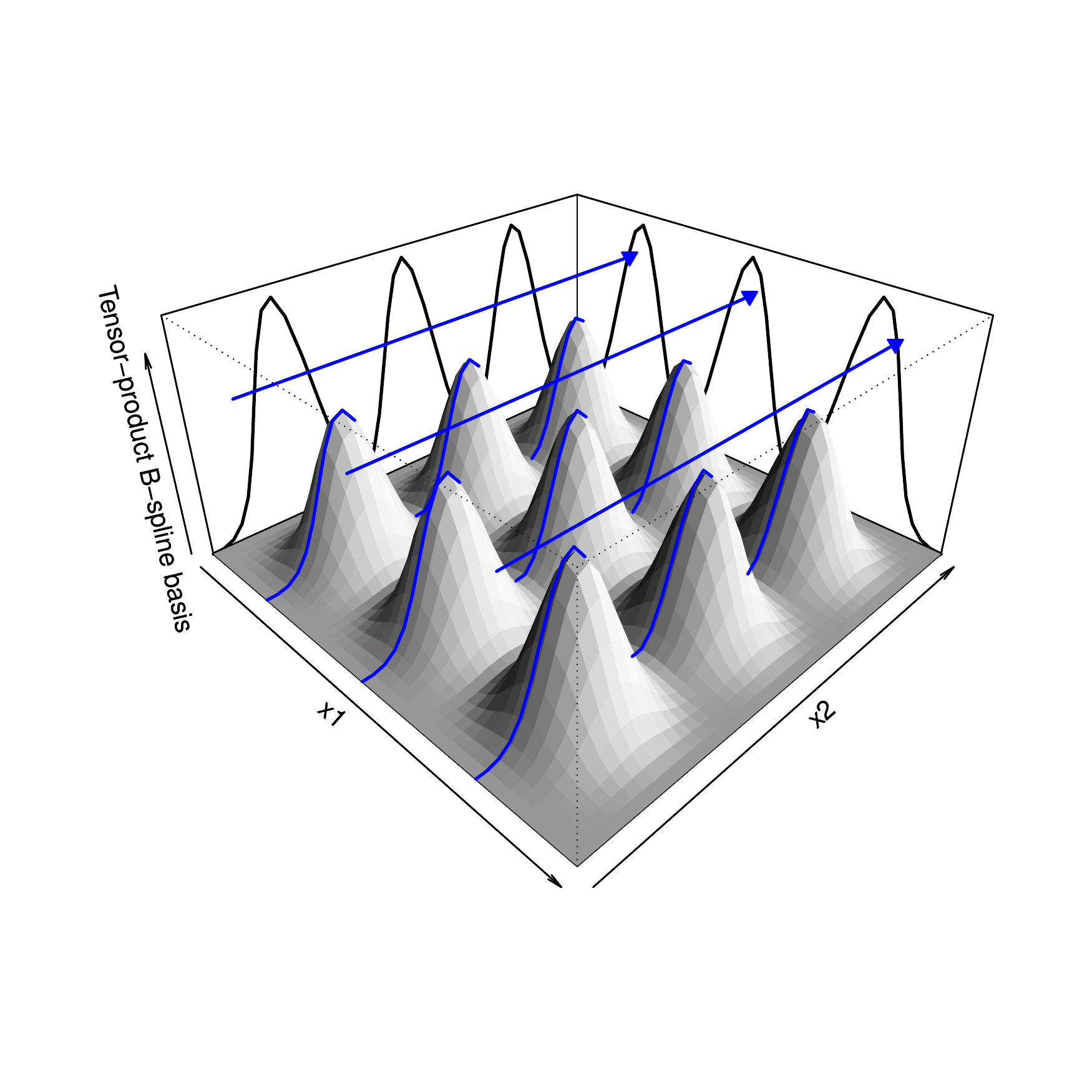}
 %\label{diff_penalty_2d_1}}
 %\subfigure[Locally adaptive penalty]{
 \includegraphics[width=7.2cm]{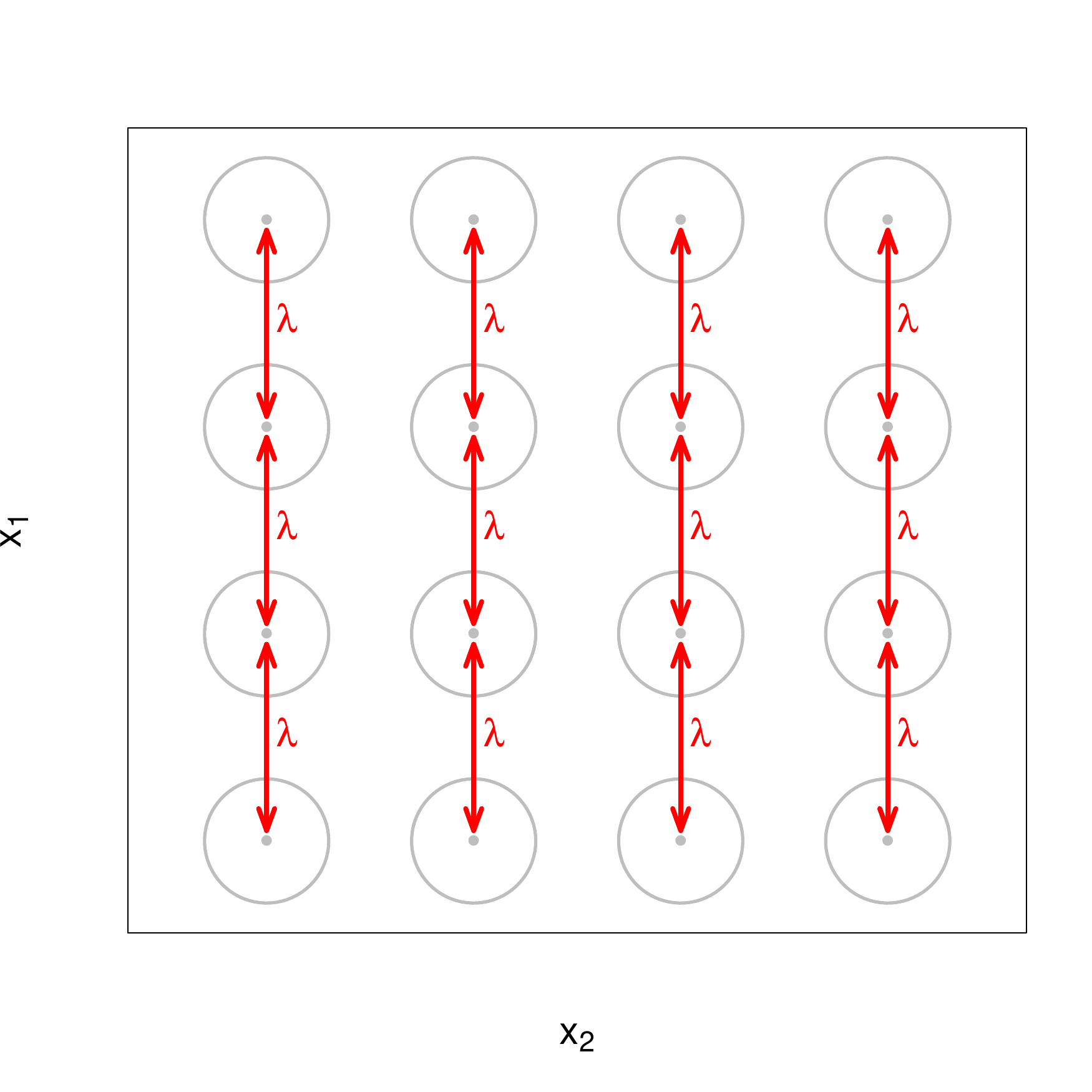}
 \includegraphics[width=7.2cm]{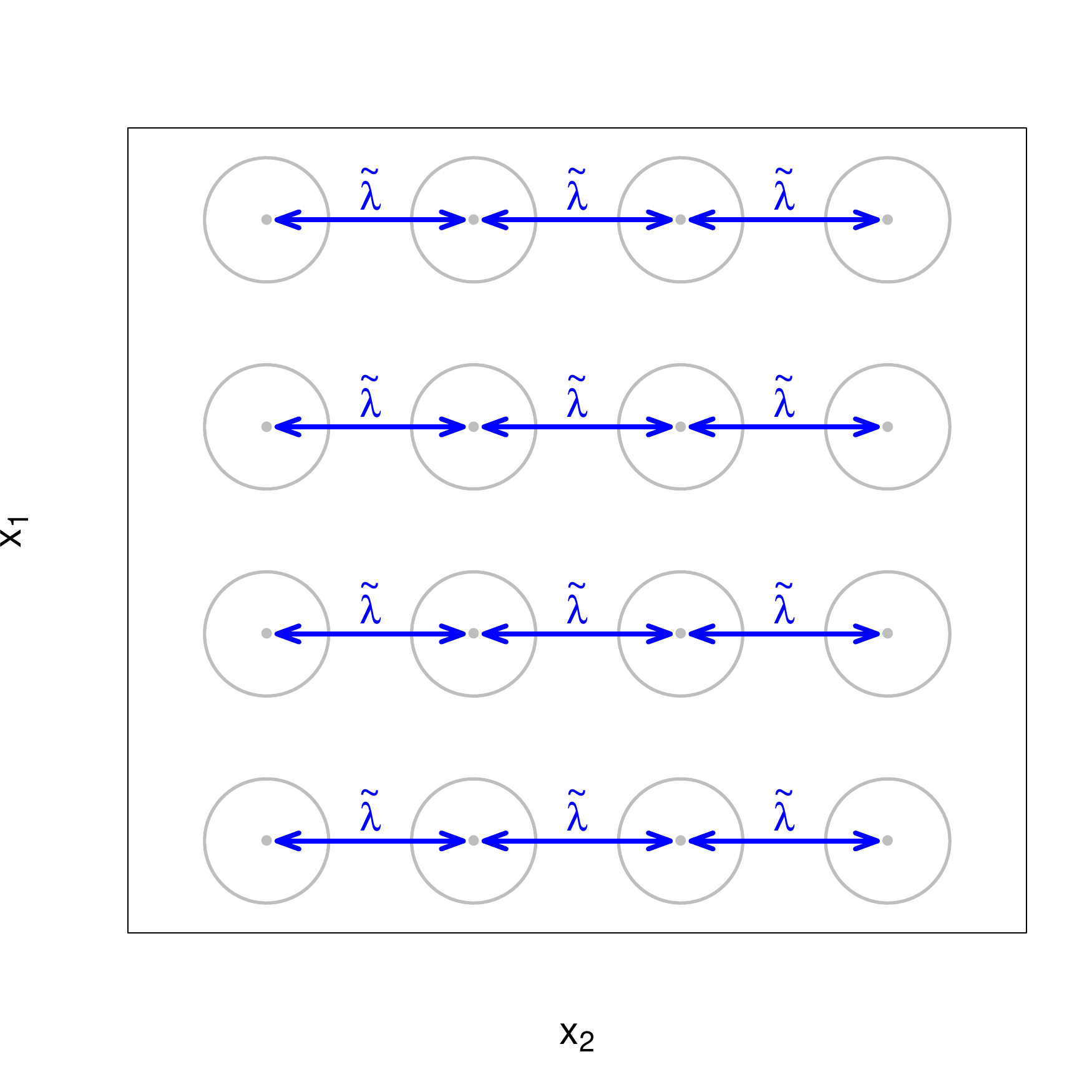}
 %\label{diff_penalty_2d_2}}
	\end{center}
	\caption{Illustration of the tensor-product of marginal B-splines basis functions ($B_{1j}\left(x_1\right)B_{2k}\left(x_2\right)$) and the anisotropic penalty (based on coefficient differences along covariates $x_1$ and $x_2$) defined in \eqref{MX:penalty_iso_2d}. The top row shows the landscape of nine cubic B-spline tensor-products -- a portion of a full basis -- and highlights why forming coefficient differences along $x_1$ and $x_2$ (i.e., on,  respectively, the columns and rows of the matrix of coefficients) ensures smoothness along the corresponding covariate. The bottom row schematically illustrates the coefficient differences (arrows) and the smoothing parameters acting on them. In both cases, red is used for covariate $x_1$ and blue for covariate $x_2$.}
	\label{MX:diff_penalty_2d}
\end{figure}

In the two-dimensional case, interest lies on the generalised model
\begin{equation}
g\left(\mu_i\right) = f\left(x_{i1}, x_{i2}\right), \;\;\;\;\; i = 1,\ldots,n,
\label{MX:PsplineM_2D}
\end{equation}
where $\boldsymbol{x}_i = \left(x_{i1}, x_{i2}\right)^{\top}$ is a two-dimensional covariate vector, and $f\left(\cdot, \cdot\right)$ is a smooth and unknown bivariate surface, defined over covariates $x_1$ and $x_2$. When it comes to extend the P-spline principles to two dimensions, we first model the bidimensional surface in terms of B-splines. This is accomplished by the tensor-product of two marginal B-splines bases, i.e., 
\[
f(x_1,x_2) = \sum_{j=1}^{d_1}\sum_{k=1}^{d_2}\theta_{jk}B_{1j}\left(x_1\right)B_{2k}\left(x_2\right),
\]
where $B_{1j}\left(\cdot\right)$ and $B_{2k}\left(\cdot\right)$ are the marginal B-spline basis functions for, respectively, $x_1$ and $x_2$. Expressed in matrix form, model \eqref{MX:PsplineM_2D} becomes
\begin{equation}
g\left(\boldsymbol{\mu}\right) = \left(\boldsymbol{B}_2\Box\boldsymbol{B}_1\right)\boldsymbol{\theta} = \boldsymbol{B}\boldsymbol{\theta},
\label{MX:PsplineM_2D_v2}
\end{equation}
where
\[
\boldsymbol{B} =\boldsymbol{B}_2\Box \boldsymbol{B}_1 = \left(\boldsymbol{B}_2\otimes\boldsymbol{1}^{\top}_{d_1}\right)\odot \left(\boldsymbol{1}_{d_2}^{\top}\otimes \boldsymbol{B}_1\right),
\] 
and $\boldsymbol{B}_1 = [b_{1;ij}]$ with $b_{1;ij} = B_{1j}(x_{1i})$, $\boldsymbol{B}_2 = [b_{s;ik}]$ with $b_{2;ik} = B_{2k}(x_{2i})$, $\boldsymbol{1}_{N}$ is a column vector of ones of length $N$, $\otimes$ denotes the Kronecker product, $\odot$ the element-wise (Hadamard) product, and $\Box$ the ``box'' product \cite[the face-splitting product or row-wise Kronecker product,][]{Slyusar99, Eilers2006}. Finally, $\boldsymbol{\mu} = \left(\mu_1, \ldots, \mu_n\right)^{\top}$ and $\boldsymbol{\theta} = \left(\theta_{11},\ldots,\theta_{d_11},\ldots, \theta_{d_1d_2}\right)^{\top}$. As for the one-dimensional case, in the two-dimensional setting smoothness is achieved by penalising coefficient differences. In particular, the anisotropic penalty in two dimensions is defined as
\begin{equation}
\boldsymbol{\theta}^{\top}\left(\lambda\left(\boldsymbol{I}_{d_2}\otimes\boldsymbol{D}_{q_1}^{\top}\boldsymbol{D}_{q_1}\right) + \widetilde{\lambda} \left(\boldsymbol{D}_{q_2}\boldsymbol{D}_{q_2}^{\top}\otimes\boldsymbol{I}_{d_1}\right)\right)\boldsymbol{\theta},
\label{MX:penalty_iso_2d}
\end{equation}
where $\boldsymbol{D}_{q_m}$ ($m = 1,2$) are difference matrices of possibly different order $q_m$, and $\lambda$ and $\widetilde{\lambda}$ are the smoothing parameters \citep{Eilers2003}. Before proceeding, it is worth seeing the vector $\boldsymbol{\theta}$ as a $(d_1 \times d_2)$ matrix of coefficients, $\boldsymbol{\Theta} = [\theta_{jk}]$; the rows and columns of $\boldsymbol{\Theta}$ correspond to the regression coefficients in the $x_1$ and $x_2$ direction, respectively. Thus, $\boldsymbol{\theta}^{\top}\left(\boldsymbol{I}_{d_2}\otimes\boldsymbol{D}_{q_1}^{\top}\boldsymbol{D}_{q_1}\right)\boldsymbol{\theta}$ forms (the sum of squares of) differences of order $q_1$ on each column of the matrix of coefficients $\boldsymbol{\Theta}$; it is thus responsible, in combination with the smoothing parameter $\lambda$, for the smoothness along covariate $x_1$. Similarly, $\boldsymbol{\theta}^{\top}\left(\boldsymbol{D}_{q_2}\boldsymbol{D}_{q_2}^{\top}\otimes\boldsymbol{I}_{d_1}\right)\boldsymbol{\theta}$ forms (the sum of squares of) differences of order $q_2$ on each row of the matrix of coefficients, controlling, jointly with $\widetilde{\lambda}$, the smoothness along covariate $x_2$. We provide Figure \ref{MX:diff_penalty_2d} to give more insights about (the reasoning behind) the anisotropic penalty just presented as well as to help presenting the adaptive penalty in two dimensions that will follow.

\begin{figure}[!h]
 \begin{center}
  \includegraphics[width=7.5cm]{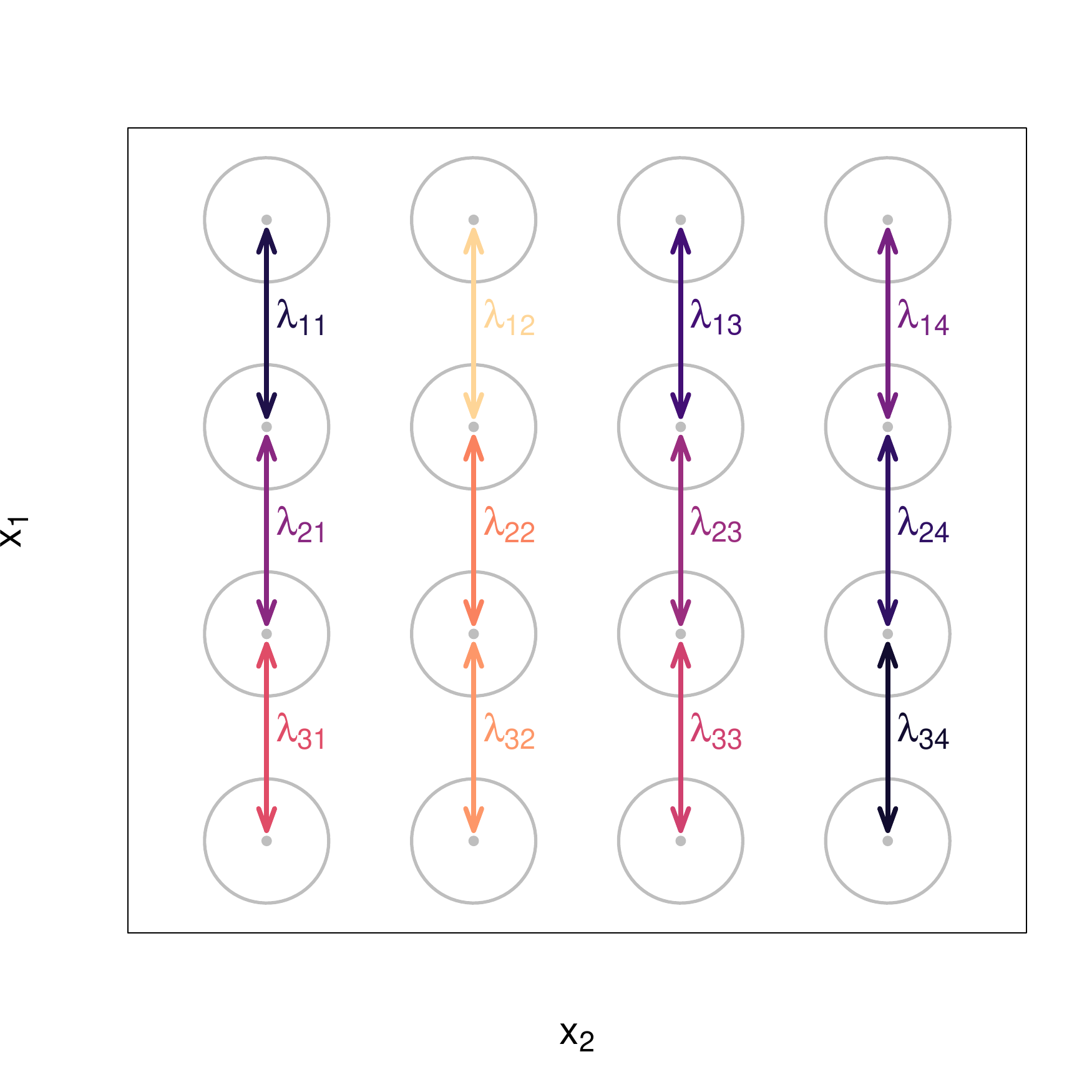}
  \includegraphics[width=7.5cm]{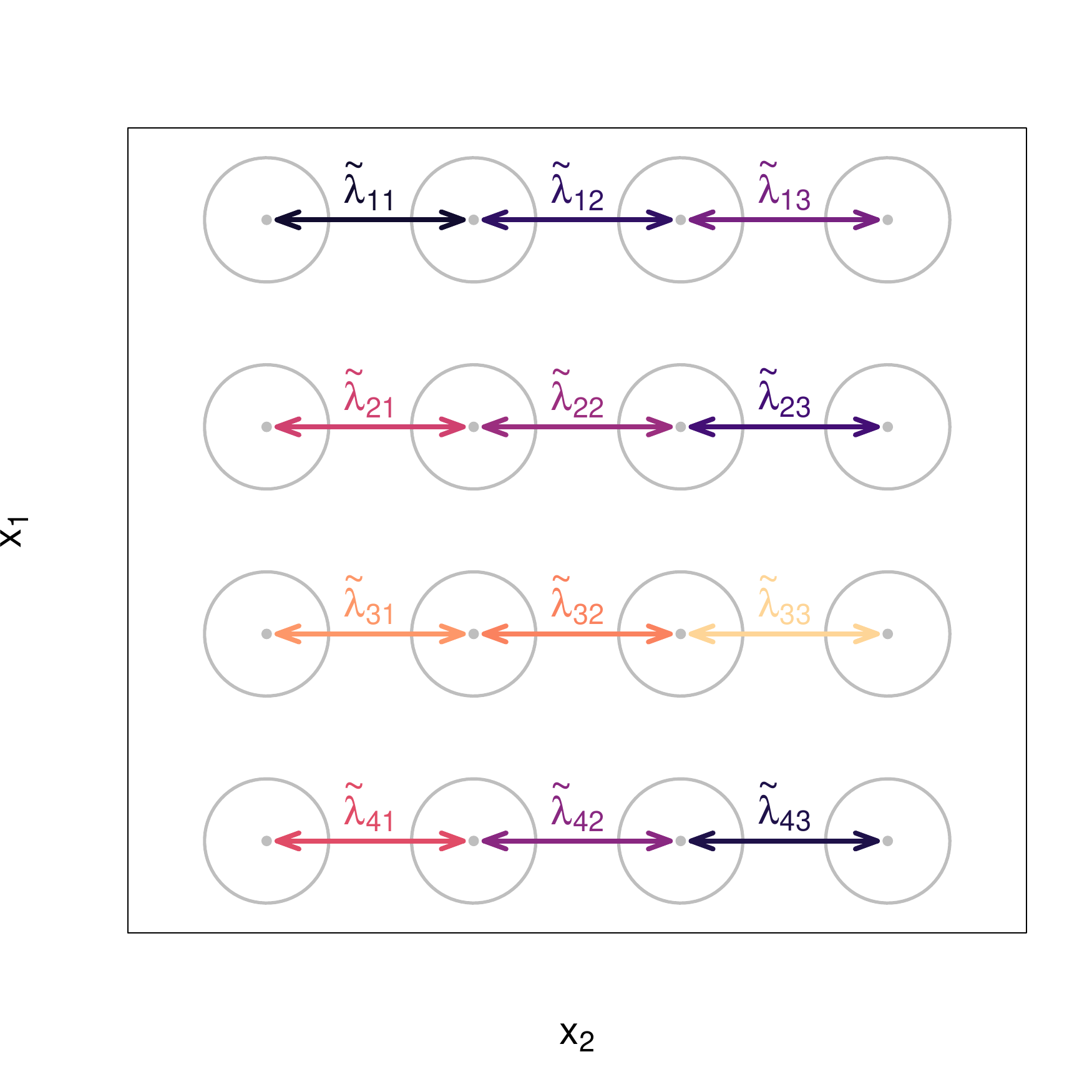}
 \end{center}
 \caption{Illustration of the adaptive penalty in two dimensions. Separately for $x_1$ (left-hand side plot) and $x_2$ (right-hand side plot), each coefficient difference is penalised by a different smoothing parameter. The arrows (and colors) schematically represent different coefficient differences jointly with the smoothing parameters acting on them ($\lambda_{uv}$ for the coefficient differences along $x_1$ and $\widetilde{\lambda}_{sp}$ for the coefficient differences along $x_2$).}
	\label{MX:diff_add_penalty_2d}
\end{figure}

By considering two different smoothing parameters $\lambda$ and $\widetilde{\lambda}$ (i.e., anisotropy), the penalty in \eqref{MX:penalty_iso_2d} permits a different amount of smoothing for $x_1$ and $x_2$. However, in some circumstances, this flexibility may not be enough for the model to capture ``local'' behaviours in the data; all coefficient differences (along $x_1$ or along $x_2$) are penalised by the same smoothing parameter ($\lambda$ or $\widetilde{\lambda}$), and thus the same amount of smoothing is assumed along each covariate. Following the same reasoning as in the one-dimensional case, we propose to overcome the (possible) lack of flexibility of penalty \eqref{MX:penalty_iso_2d} by considering a different smoothing parameter for each coefficient difference, and we do that separately for the coefficient differences along $x_1$ and along $x_2$. The idea is graphically exemplified in Figure \ref{MX:diff_add_penalty_2d}, which shows clearly that our approach gives rise to two matrices of smoothing parameters, $\boldsymbol{\Lambda} = [\lambda_{uv}]$ of dimension $\left(d_1 - q_1\right)\times d_2$, and $\widetilde{\boldsymbol{\Lambda}} = [\widetilde{\lambda}_{sp}]$ of dimension $d_1 \times \left(d_2 - q_2\right)$. Recall that $d_m$ is the dimension of the marginal B-splines bases and $q_m$ is the penalty order ($m = 1,2$). In particular, the smoothing parameters in $\boldsymbol{\Lambda}$ will act on the coefficient differences along $x_1$, and will then control the amount of smoothing along that covariate, but permitting it to vary locally. The same applies to $\widetilde{\boldsymbol{\Lambda}}$, which (adaptively) controls the amount of smoothing along $x_2$. With this in mind, our two-dimensional anisotropic adaptive penalty adopts the following form
\begin{equation}
\boldsymbol{\theta}^{\top}\left(\left(\boldsymbol{I}_{d_2}\otimes\boldsymbol{D}_{q_1}\right)^{\top}\mbox{diag}(\boldsymbol{\lambda})\left(\boldsymbol{I}_{d_2}\otimes\boldsymbol{D}_{q_1}\right) + \left(\boldsymbol{D}_{q_2}\otimes\boldsymbol{I}_{d_1}\right)^{\top}\mbox{diag}(\widetilde{\boldsymbol{\lambda}})\left(\boldsymbol{D}_{q_2}\otimes\boldsymbol{I}_{d_1}\right)\right)\boldsymbol{\theta},
\label{MX:add_penalty_2d}
\end{equation} 
where $\boldsymbol{\lambda} = \mbox{vec}\left(\boldsymbol{\Lambda}\right)$ and $\widetilde{\boldsymbol{\lambda}} = \mbox{vec}\left(\widetilde{\boldsymbol{\Lambda}}\right)$, and $\mbox{vec}(\boldsymbol{M})$ denotes the vectorisation of matrix $\boldsymbol{M}$. A possible drawback of the adaptive penalty defined in \eqref{MX:add_penalty_2d} is the number of smoothing parameters involved, which equals to $\left(d_1 - q_1\right)\times d_2 + d_1 \times \left(d_2 - q_2\right)$; i.e., the total number of coefficient differences along $x_1$ and $x_2$. This may give rise, in addition to undersmoothing and unstable computations, to prohibitively long computing times. We propose to reduce the complexity of the multidimensional adaptive penalty in \eqref{MX:add_penalty_2d} through a reduction on the number of smoothing parameters. In like manner to the one-dimensional case presented in Section \ref{MX:adapt_pen_1D}, this is done by considering a smoothed (and smaller) version of them. The underlying assumption is that smoothing parameters that are spatially proximate are more likely to be similar than those farther apart. Before proceeding, note that we now have two matrices of smoothing parameters $\boldsymbol{\Lambda}$ and $\widetilde{\boldsymbol{\Lambda}}$ (see also Figure \ref{MX:diff_add_penalty_2d}). It seems then reasonable to smooth them using the tensor-product of marginal B-spline bases, and we do it separately for $\boldsymbol{\Lambda}$ and $\widetilde{\boldsymbol{\Lambda}}$. Taking the advantage of the ``data'' (smoothing parameters) being in an array structure, we write 
\begin{align}	
\boldsymbol{\lambda} = \big(\boldsymbol{\Psi}_{2}\otimes\boldsymbol{\Psi}_{1}\big)\boldsymbol{\xi}, \label{MX:smooth_ad_2d_1} \\ 
\widetilde{\boldsymbol{\lambda}} = \big(\widetilde{\boldsymbol{\Psi}}_{2}\otimes\widetilde{\boldsymbol{\Psi}}_{1}\big)\widetilde{\boldsymbol{\xi}}, \label{MX:smooth_ad_2d_2}
\end{align}
where $\boldsymbol{\xi} = \left(\xi_{1},\ldots,\xi_{p_{11}p_{12}}\right)^{\top}$ and $\widetilde{\boldsymbol{\xi}} = \left(\widetilde{\xi}_{1},\ldots,\widetilde{\xi}_{p_{21}p_{22}}\right)^{\top}$ are the new vectors of smoothing parameters, and $\boldsymbol{\Psi}_{1}^{(d_1-q_1)\times p_{11}}$, $\boldsymbol{\Psi}_{2}^{d_2\times p_{12}}$, $\widetilde{\boldsymbol{\Psi}}_{1}^{d_1\times p_{21}}$ and $\widetilde{\boldsymbol{\Psi}}_{2}^{(d_2-q_2)\times p_{22}}$ are B-spline design matrices (the super-indices indicate their dimension). In particular, these matrices are constructed as follows
\begin{equation}
\begin{aligned}
\boldsymbol{\Psi}_1 & = \left(
\begin{array}{ccc}
\psi_{11}(1) & \ldots  & \psi_{1p_{11}}(1)\\
\vdots & \ddots & \vdots\\
\psi_{11}(d_1 - q_1) & \ldots  & \psi_{1p_{11}}(d_1 - q_1)
\end{array}
\right),\\[12pt]
\widetilde{\boldsymbol{\Psi}}_1 & = \left(
\begin{array}{ccc}
\widetilde{\psi}_{11}(1) & \ldots  & \widetilde{\psi}_{1p_{21}}(1)\\
\vdots & \ddots & \vdots\\
\widetilde{\psi}_{11}(d_1) & \ldots  & \widetilde{\psi}_{1p_{21}}(d_1)
\end{array}
\right),
\end{aligned}
\begin{aligned}
\boldsymbol{\Psi}_2 & = \left(
\begin{array}{ccc}
\psi_{21}(1) & \ldots  & \psi_{2p_{12}}(1)\\
\vdots & \ddots & \vdots\\
\psi_{21}(d_2) & \ldots  & \psi_{2p_{12}}(d_2)
\end{array}
\right),\\[12pt]
\widetilde{\boldsymbol{\Psi}}_2 & = \left(
\begin{array}{ccc}
\widetilde{\psi}_{21}(1) & \ldots  & \widetilde{\psi}_{2p_{22}}(1)\\
\vdots & \ddots & \vdots\\
\widetilde{\psi}_{21}(d_2 - q_2) & \ldots  & \widetilde{\psi}_{2p_{22}}(d_2 - q_2)
\end{array}
\right).
\end{aligned}
\label{2d_matrices}
\end{equation}
Plugging-in the right-hand side of equations \eqref{MX:smooth_ad_2d_1} and \eqref{MX:smooth_ad_2d_2} into \eqref{MX:add_penalty_2d}, and after some algebraic operations, we obtain our proposal for the adaptive penalty in two dimensions
\begin{align}
\boldsymbol{\theta}^{\top}\Biggl(&\sum_{u = 1}^{p_{11}p_{12}}\xi_{u}\left(\mathbf{I}_{d_2}\otimes\boldsymbol{D}_{q_1}\right)^{\top}\mbox{diag}\left(\boldsymbol{\psi}_{u}\right)\left(\mathbf{I}_{d_2}\otimes\boldsymbol{D}_{q_1}\right) + \nonumber \\
&\sum_{s = 1}^{p_{21}p_{22}}\widetilde{\xi}_{s}\left(\boldsymbol{D}_{q_2}\otimes\mathbf{I}_{d_1}\right)^{\top}\mbox{diag}\left(\widetilde{\boldsymbol{\psi}}_{s}\right)\left(\boldsymbol{D}_{q_2}\otimes\mathbf{I}_{d_1}\right)\Biggr)\boldsymbol{\theta},
\label{MX:add_penalty_2d_reduced}
\end{align}
where $\boldsymbol{\psi}_{u}$ and $\widetilde{\boldsymbol{\psi}}_{s}$ denote, respectively, the columns $u$ and $s$ of $\boldsymbol{\Psi} = \boldsymbol{\Psi}_{2}\otimes\boldsymbol{\Psi}_{1}$ (see (\ref{MX:smooth_ad_2d_1})) and $\widetilde{\boldsymbol{\Psi}} = \widetilde{\boldsymbol{\Psi}}_{2}\otimes\widetilde{\boldsymbol{\Psi}}_{1}$ (see (\ref{MX:smooth_ad_2d_2})).
%of the matrices of smoothing parameters $\boldsymbol{\Lambda}$ and $\widetilde{\boldsymbol{\Lambda}}$
%; the number of smoothing parameters to be estimated can be very large
%This figure graphically illustrates the reasoning behind the anisotropic penalty in two dimensions just described.
\subsubsection{Simplifications}\label{simplifications}
The two-dimensional adaptive penalty presented in the previous section (expression (\ref{MX:add_penalty_2d_reduced})) is the most general one: A different smoothing parameter is assumed for each coefficient difference along both $x_1$ and $x_2$. However, several simplifications may be made according to the data at hand. 

\begin{description}
\item[S. I] The most obvious simplification is to consider an adaptive penalty for one of the covariates (say $x_1$) in combination with a traditional (non-adaptive) penalty for $x_2$. This would correspond to the left-hand side plot of Figure \ref{MX:diff_add_penalty_2d} and the right-hand side plot of Figure \ref{MX:diff_penalty_2d}, respectively. In such case, it is easy to show that the simplified two-dimensional adaptive penalty adopts the form
\begin{equation}
\boldsymbol{\theta}^{\top}\left(\sum_{u = 1}^{p_{11}p_{12}}\xi_{u}\left(\mathbf{I}_{d_2}\otimes\boldsymbol{D}_{q_1}\right)^{\top}\mbox{diag}\left(\boldsymbol{\psi}_{u}\right)\left(\mathbf{I}_{d_2}\otimes\boldsymbol{D}_{q_1}\right) + \widetilde{\lambda}\left(\boldsymbol{D}_{q_2}\boldsymbol{D}_{q_2}^{\top}\otimes\boldsymbol{I}_{d_1}\right)\right)\boldsymbol{\theta},
\label{MX:add_penalty_2d_reduced_case_I}
\end{equation}
with $\boldsymbol{\psi}_{u}$ defined as in (\ref{MX:add_penalty_2d_reduced}).
\end{description}

\begin{figure}[!h]
 \begin{center}
  \subfigure[Simplification II]{\includegraphics[width=7.5cm]{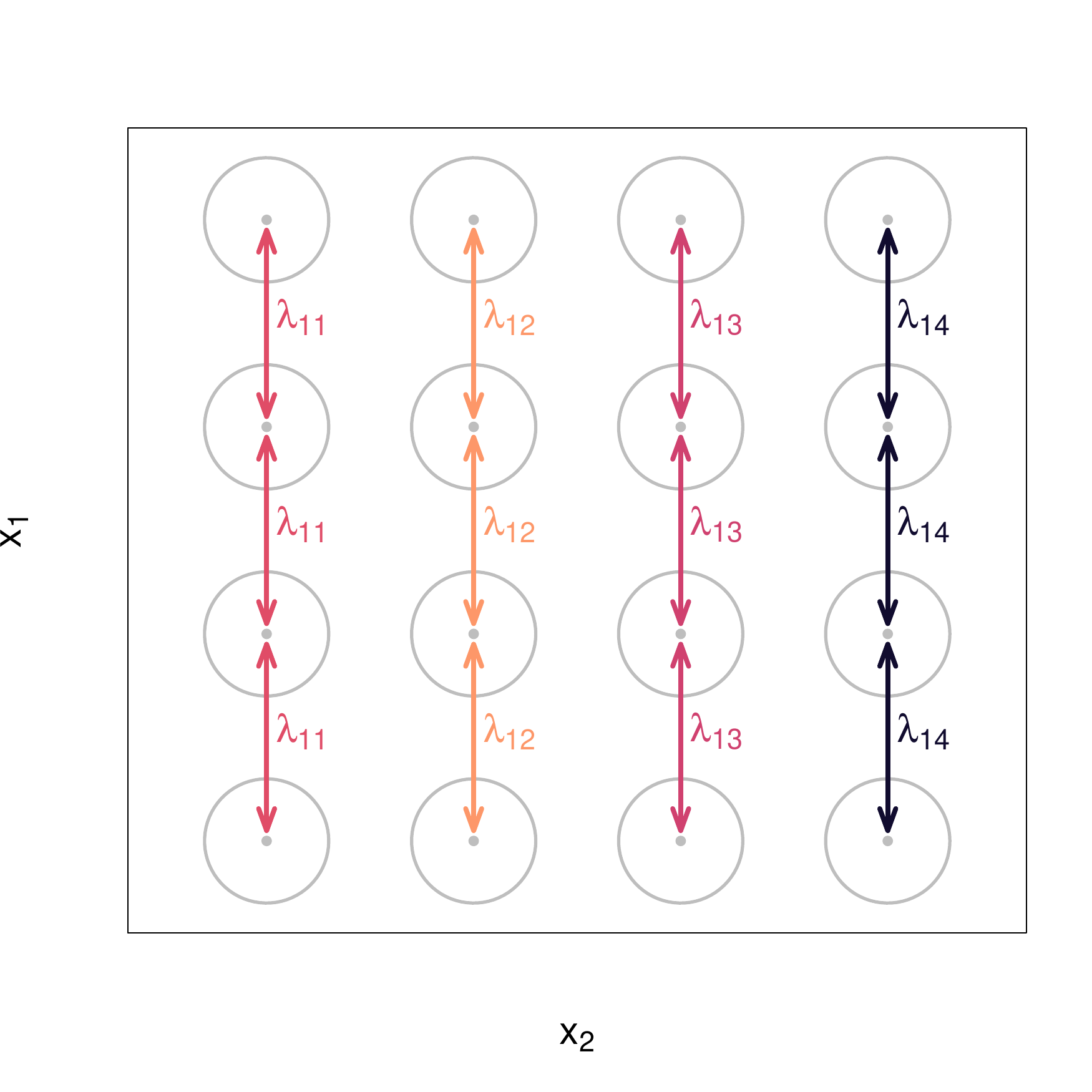}\label{MX:diff_add_penalty_2d_simpl_A}}
  \subfigure[Simplification III]{\includegraphics[width=7.5cm]{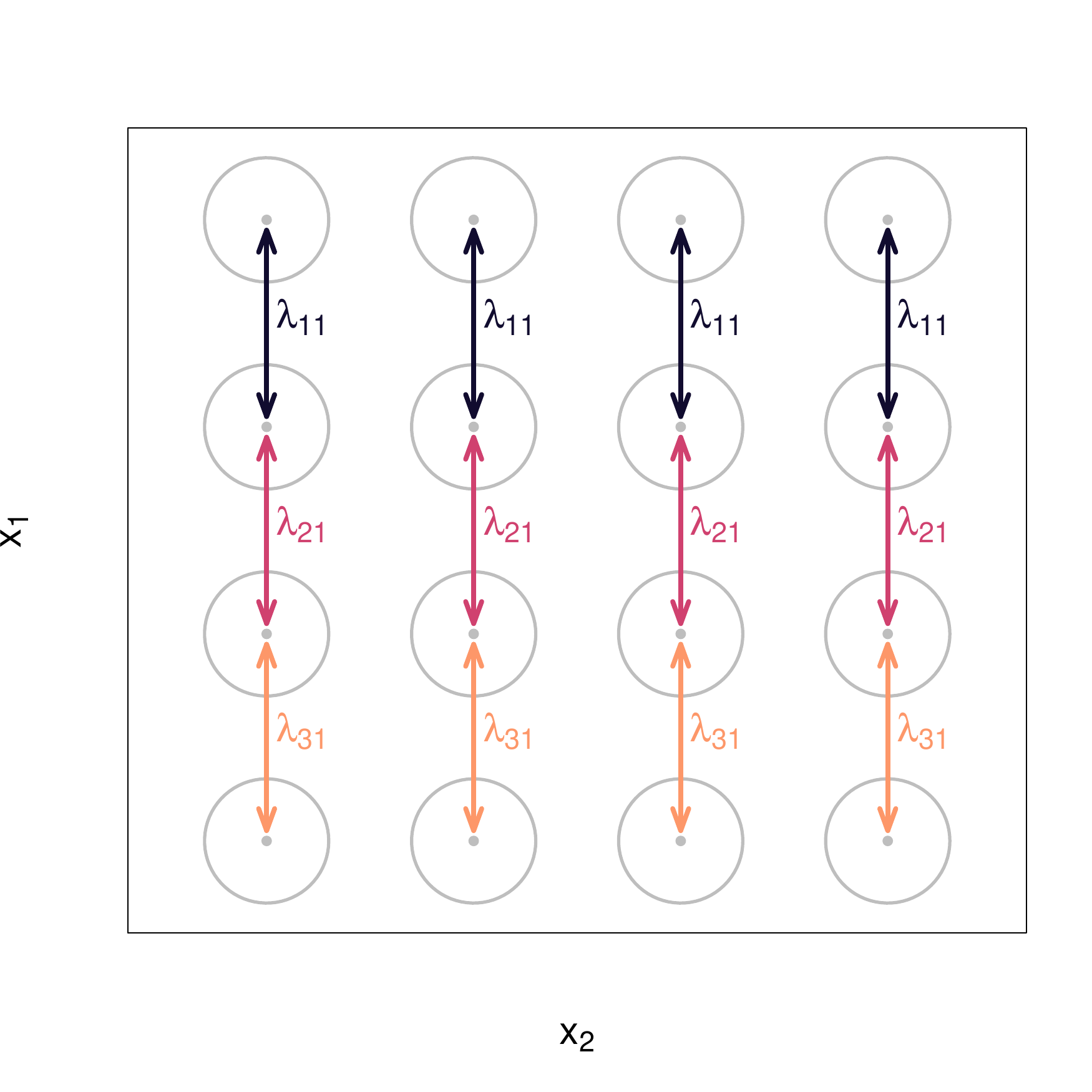}\label{MX:diff_add_penalty_2d_simpl_B}}
 \end{center}
 \caption{Illustration of the simplifications on the full adaptive penalty in two dimensions along $x_1$. The arrows schematically represent different coefficient differences jointly with the smoothing parameters acting on them. Simplification II corresponds to the situation in which it is assumed that the smoothness along $x_1$ changes according to (needs to adapt to) the values of $x_2$. Simplification III describes the case when adaptive smoothing is needed along $x_1$, but it does not change across $x_2$ values.}
\end{figure}

Further interesting simplifications can also be made to the full adaptive penalty along each covariate. For clarity, these will be described for $x_1$, but an analogous reasoning applies to $x_2$. 
\begin{description}
\item[S. II] One possible simplification is to assume that no adaptive smoothing is needed along $x_1$, but that the smoothness along that covariate needs to change according to (needs to adapt to) the values of $x_2$. This is illustrated in Figure \ref{MX:diff_add_penalty_2d_simpl_A}. Note that this equals to assume a unique smoothing parameter for all coefficient differences formed along each column of the matrix $\boldsymbol{\Theta}$, but with that smoothing parameter being different for each column. In this case, equation \eqref{MX:smooth_ad_2d_1} would reduce to
\[
\boldsymbol{\lambda} = \big(\boldsymbol{\Psi}_{2} \otimes \boldsymbol{1}_{d_1-q_1}\big)\boldsymbol{\xi},
\]
where $\boldsymbol{\lambda} = \left(\lambda_{11}, \ldots, \lambda_{1d_2}\right)^{\top}  \otimes \boldsymbol{1}_{d_1-q_1}$, $\boldsymbol{\xi} = \left(\xi_1, \ldots, \xi_{p_{12}}\right)^{\top}$. We emphasise that the difference with respect to the full case is that $\boldsymbol{\Psi}_{1}$ in \eqref{MX:smooth_ad_2d_1} is replaced by a column-vector of ones (it also implies that the length of vector $\boldsymbol{\xi}$ is lower than in the full case). The two-dimensional adaptive penalty (assuming a fully adaptive one for $x_2$) thus becomes
\begin{align}
\boldsymbol{\theta}^{\top}\Biggl(& \sum_{u = 1}^{p_{12}}\xi_{u}\left(\mathbf{I}_{d_2}\otimes\boldsymbol{D}_{q_1}\right)^{\top}\mbox{diag}\left(\boldsymbol{\psi}^{\text{II}}_{u}\right)\left(\mathbf{I}_{d_2}\otimes\boldsymbol{D}_{q_1}\right) + \nonumber \\
& \sum_{s = 1}^{p_{21}p_{22}}\widetilde{\xi}_{s}\left(\boldsymbol{D}_{q_2}\otimes\mathbf{I}_{d_1}\right)^{\top}\mbox{diag}\left(\widetilde{\boldsymbol{\psi}}_{s}\right)\left(\boldsymbol{D}_{q_2}\otimes\mathbf{I}_{d_1}\right)\Biggr)\boldsymbol{\theta},
\label{MX:add_penalty_2d_reduced_case_II}
\end{align}
where $\boldsymbol{\psi}^{\text{II}}_{u}$ and $\widetilde{\boldsymbol{\psi}}_{s}$ denote, respectively, the columns $u$ and $s$ of $\boldsymbol{\Psi}^{\text{II}} = \boldsymbol{\Psi}_{2} \otimes \boldsymbol{1}_{d_1-q_1}$ and $\widetilde{\boldsymbol{\Psi}} = \widetilde{\boldsymbol{\Psi}}_{2}\otimes\widetilde{\boldsymbol{\Psi}}_{1}$, with $\boldsymbol{\Psi}_{2}$, $\widetilde{\boldsymbol{\Psi}}_{2}$ and $\widetilde{\boldsymbol{\Psi}}_{1}$ defined as in (\ref{2d_matrices}).
\item[S. III] Another simplification is to assume that adaptive smoothing is needed along $x_1$, but that it does not change across $x_2$ values. The idea is presented in Figure \ref{MX:diff_add_penalty_2d_simpl_B}. Note that here we have that
\[
\boldsymbol{\lambda} = \big(\boldsymbol{1}_{d_2} \otimes \boldsymbol{\Psi}_{1} \big)\boldsymbol{\xi},
\]
where $\boldsymbol{\lambda} =  \boldsymbol{1}_{d_2} \otimes \left(\lambda_{11}, \ldots, \lambda_{(d_1-q_1)1}\right)^{\top}$, $\boldsymbol{\xi} = \left(\xi_1, \ldots, \xi_{p_{11}}\right)^{\top}$, and thus the adaptive penalty is
\begin{align}
\boldsymbol{\theta}^{\top}\Biggl(&\sum_{u = 1}^{p_{11}}\xi_{u}\left(\mathbf{I}_{d_2}\otimes\boldsymbol{D}_{q_1}\right)^{\top}\mbox{diag}\left(\boldsymbol{\psi}^{\text{III}}_{u}\right)\left(\mathbf{I}_{d_2}\otimes\boldsymbol{D}_{q_1}\right) + \nonumber \\
&\sum_{s = 1}^{p_{21}p_{22}}\widetilde{\xi}_{s}\left(\boldsymbol{D}_{q_2}\otimes\mathbf{I}_{d_1}\right)^{\top}\mbox{diag}\left(\widetilde{\boldsymbol{\psi}}_{s}\right)\left(\boldsymbol{D}_{q_2}\otimes\mathbf{I}_{d_1}\right)\Biggr)\boldsymbol{\theta},
\label{MX:add_penalty_2d_reduced_case_III}
\end{align}
where $\boldsymbol{\psi}^{\text{III}}_{s}$ and $\widetilde{\boldsymbol{\psi}}_{u}$ denote, respectively, the columns $s$ and $u$ of $\boldsymbol{\Psi}^{\text{III}} = \boldsymbol{1}_{d_2} \otimes \boldsymbol{\Psi}_{1}$ and $\widetilde{\boldsymbol{\Psi}} = \widetilde{\boldsymbol{\Psi}}_{2}\otimes\widetilde{\boldsymbol{\Psi}}_{1}$, with $\boldsymbol{\Psi}_{1}$, $\widetilde{\boldsymbol{\Psi}}_{2}$ and $\widetilde{\boldsymbol{\Psi}}_{1}$ defined as in (\ref{2d_matrices}).
\end{description}
\subsection{Extension to Three Dimensions}
With all the ingredients that have been presented in the previous section, the extension of the adaptive penalty (and its simplifications) to the three-dimensional case is straightforward. We now focus on
\begin{equation}
g\left(\mu_i\right) = f\left(x_{i1}, x_{i2}, x_{i3}\right), \;\;\;\;\; i = 1,\ldots,n,
\label{MX:PsplineM_3D}
\end{equation}
where $\boldsymbol{x}_i = \left(x_{i1}, x_{i2}, x_{i3}\right)^{\top}$ is a three-dimensional covariate vector. Again, $f\left(\cdot, \cdot, \cdot\right)$ is approximated by the tensor-product of three marginal B-splines bases, leading to the higher order box product
\[
g\left(\boldsymbol{\mu}\right) = \left(\boldsymbol{B}_3\Box\boldsymbol{B}_2\Box\boldsymbol{B}_1\right)\boldsymbol{\theta} = \boldsymbol{B}\boldsymbol{\theta},
\]
where $\boldsymbol{B}_1^{n \times d_1}$, $\boldsymbol{B}_2^{n \times d_2}$ and $\boldsymbol{B}_3^{n \times d_3}$ are B-spline design matrices for, respectively, $x_1$, $x_2$ and $x_3$, and smoothness is achieved by penalising coefficient differences along each covariate, i.e., the anisotropic penalty in three dimensions is
\begin{equation*}
\boldsymbol{\theta}^{\top}\left(\lambda_1\left(\boldsymbol{I}_{d_3}\otimes\boldsymbol{I}_{d_2}\otimes\boldsymbol{D}_{q_1}^{\top}\boldsymbol{D}_{q_1}\right) + \lambda_2 \left(\boldsymbol{I}_{d_3}\otimes\boldsymbol{D}_{q_2}\boldsymbol{D}_{q_2}^{\top}\otimes\boldsymbol{I}_{d_1}\right) + \lambda_3 \left(\boldsymbol{D}_{q_3}\boldsymbol{D}_{q_3}^{\top}\otimes\boldsymbol{I}_{d_2}\otimes \boldsymbol{I}_{d_1}\right)\right)\boldsymbol{\theta}.
\label{MX:penalty_iso_3d}
\end{equation*}
%Note that there are three smoothing parameters $\lambda_1$, $\lambda_2$ and $\lambda_3$. We now extend the above anisotropic penalty by first considering a different smoothing parameter for each coefficient difference (this is done separately for $x_1$, $x_2$ and $x_3$), and then smoothing (separately) the three three-dimensional arrays of smoothing parameters that this approach gives rise, i.e.,
Note that there are three smoothing parameters $\lambda_1$, $\lambda_2$ and $\lambda_3$. We now extend the above anisotropic penalty by considering a different smoothing parameter for each coefficient difference; this is done separately for $x_1$, $x_2$ and $x_3$. In this case, this will give rise to three (one for each covariate) three-dimensional arrays of smoothing parameters of dimensions $(d_1 - q_1) \times d_2 \times d_3$, $d_1 \times (d_2 - q_2) \times d_3$ and $d_1 \times d_2 \times (d_3 - q_3)$ for, respectively $x_1$, $x_2$ and $x_3$. Once again, we smooth them to reduce their dimensions using B-splines tensor-products, i.e.,
\begin{align}	
\boldsymbol{\lambda}_1^{(d_1 - q_1) d_2  d_3  } & = \left(\boldsymbol{\Psi}_{13}^{d_3 \times p_{13}}\otimes\boldsymbol{\Psi}_{12}^{d_2 \times p_{12}}\otimes\boldsymbol{\Psi}_{11}^{(d_1-q_1) \times p_{11}}\right)\boldsymbol{\xi}_1^{p_{11}  p_{12}  p_{13}},\label{MX:3D_smoothing_parameters_1}\\
\boldsymbol{\lambda}_2^{d_1 (d_2 - q_2) d_3} & = \left(\boldsymbol{\Psi}_{23}^{d_3 \times p_{23}}\otimes\boldsymbol{\Psi}_{22}^{(d_2 - q_2) \times p_{22}}\otimes\boldsymbol{\Psi}_{21}^{d_1 \times p_{21}}\right)\boldsymbol{\xi}_2^{p_{21}  p_{22}  p_{23}},\label{MX:3D_smoothing_parameters_2}\\
\boldsymbol{\lambda}_3^{d_1 d_2 (d_3 - q_3)} & = \left(\boldsymbol{\Psi}_{33}^{(d_3 - q_3) \times p_{33}}\otimes\boldsymbol{\Psi}_{32}^{d_2 \times p_{32}}\otimes\boldsymbol{\Psi}_{31}^{d_3 \times p_{31}}\right)\boldsymbol{\xi}_3^{p_{31}  p_{32}  p_{33}},
\label{MX:3D_smoothing_parameters_3}
\end{align} 
where $\boldsymbol{\Psi}_{mw}$ ($m,w = 1,2,3$) are B-spline design matrices constructed in a similar fashion as presented in (\ref{2d_matrices}). The full adaptive penalty in three dimensions is thus expressed as
\begin{align}
\boldsymbol{\theta}^{\top}\Biggl(&\sum_{u = 1}^{p_{11}p_{12}p_{13}}\xi_{1u}\left(\mathbf{I}_{d_3}\otimes\mathbf{I}_{d_2}\otimes\boldsymbol{D}_{q_1}\right)^{\top}\mbox{diag}\left(\boldsymbol{\psi}_{1,u}\right)\left(\mathbf{I}_{d_3}\otimes\mathbf{I}_{d_2}\otimes\boldsymbol{D}_{q_1}\right) + \nonumber\\
&\sum_{s = 1}^{p_{21}p_{22}p_{23}}\xi_{2s}\left(\mathbf{I}_{d_3}\otimes\boldsymbol{D}_{q_2}\otimes\mathbf{I}_{d_1}\right)^{\top}\mbox{diag}\left(\boldsymbol{\psi}_{2,s}\right)\left(\mathbf{I}_{d_3}\otimes\boldsymbol{D}_{q_2}\otimes\mathbf{I}_{d_1}\right) + \label{MX:add_penalty_3d_reduced} \\
&\sum_{v = 1}^{p_{31}p_{32}p_{33}}\xi_{3v}\left(\boldsymbol{D}_{q_3}\otimes\mathbf{I}_{d_2}\otimes\mathbf{I}_{d_1}\right)^{\top}\mbox{diag}\left(\boldsymbol{\psi}_{3,v}\right)\left(\boldsymbol{D}_{q_3}\otimes\mathbf{I}_{d_2}\otimes\mathbf{I}_{d_1}\right)\Biggr)\boldsymbol{\theta},\nonumber
\end{align}
where $\boldsymbol{\psi}_{m,l}$ denotes the column $l$ of $\boldsymbol{\Psi}_m = \boldsymbol{\Psi}_{m3}\otimes\boldsymbol{\Psi}_{m2}\otimes\boldsymbol{\Psi}_{m1}$ ($m = 1,2,3$). Similar simplifications to those discussed for the adaptive penalty in two dimensions (though more difficult to visualise) can be obtained by appropriately modifying equations \eqref{MX:3D_smoothing_parameters_1} -- \eqref{MX:3D_smoothing_parameters_3}, i.e., by replacing (some of) the $\boldsymbol{\Psi}_{mw}$ matrices by (appropriately sized) column vectors of ones. 

\section{Estimation and Computational Aspects}\label{Estimation}
To estimate models (\ref{MX:PsplineM_2D}) and (\ref{MX:PsplineM_3D}) subject to, respectively, the adaptive penalties defined in (\ref{MX:add_penalty_2d_reduced}) (or the simplifications presented in (\ref{MX:add_penalty_2d_reduced_case_I}) -- (\ref{MX:add_penalty_2d_reduced_case_III})) and~(\ref{MX:add_penalty_3d_reduced}), we adopt the equivalence between P-splines and generalised linear mixed models \citep[e.g., ][]{Currie2002, Wand2003}. In particular, we follow the proposal by \cite{Lee2010} and \cite{Lee2011} that deals with the multidimensional case. %We recall that in this approach, the smooth functions are treated as sums of fixed and random components, and the smoothing parameters are replaced by ratios of variances which are estimated by restricted maximum likelihood (REML). The key is aiming to decompose the model into the unpenalised and the penalised part. The consequence of this decomposition is that the penalty matrix of the reparametrised P-spline model is of full rank, and so is the precision matrix of the corresponding mixed model.
For simplicity, results are presented for the full two-dimensional adaptive P-spline model (a similar reasoning can be followed for the simplifications), and expressions for the three-dimensional case are relegated to the Appendix.

Let $\boldsymbol{D}^{\top}_{q_m}\boldsymbol{D}_{q_m} = \boldsymbol{U}_{m}\boldsymbol{\Sigma}_{m}\boldsymbol{U}_{m}^{\top}$ be the eigenvalue decomposition (EVD) of $\boldsymbol{D}_{q_m}^{\top}\boldsymbol{D}_{q_m}$ ($m = 1,2$). Here $\boldsymbol{U}_m$ denotes the matrix of eigenvectors and $\boldsymbol{\Sigma}_m$ the diagonal matrix of eigenvalues. Let us also denote by $\boldsymbol{U}_{m+}$ ($\boldsymbol{\Sigma}_{m+}$) and $\boldsymbol{U}_{m0}$ ($\boldsymbol{\Sigma}_{m0}$) the sub-matrices corresponding to the non-zero and zero eigenvalues, respectively. We now construct the transformation matrix as follows
\[
\boldsymbol{T} = [\boldsymbol{U}_{20}\otimes\boldsymbol{U}_{10}\mid\boldsymbol{U}_{20}\otimes\boldsymbol{U}_{1+}\mid\boldsymbol{U}_{2+}\otimes\boldsymbol{U}_{10}\mid\boldsymbol{U}_{2+}\otimes\boldsymbol{U}_{1+}],
\]
and define
\begin{align*}
\boldsymbol{T}_{0} & = [\boldsymbol{U}_{20}\otimes\boldsymbol{U}_{10}],\\
\boldsymbol{T}_{+} & = [\boldsymbol{U}_{20}\otimes\boldsymbol{U}_{1+}\mid\boldsymbol{U}_{2+}\otimes\boldsymbol{U}_{10}\mid\boldsymbol{U}_{2+}\otimes\boldsymbol{U}_{1+}].
\end{align*}
Model (\ref{MX:PsplineM_2D_v2}) is then re-expressed as
\begin{equation}
g\left(\boldsymbol{\mu}\right) = \left(\boldsymbol{B}_2\Box\boldsymbol{B}_1\right)\boldsymbol{\theta} = \boldsymbol{B}\boldsymbol{\theta} = \boldsymbol{B}\boldsymbol{T}\boldsymbol{T}^{\top}\boldsymbol{\theta} = \boldsymbol{X}\boldsymbol{\beta} + \boldsymbol{Z}\boldsymbol{\alpha},
\label{MX:2DMix}
\end{equation}
where
\begin{align}
\boldsymbol{X} & = \boldsymbol{B}\boldsymbol{T}_{0} = \left[\boldsymbol{B}_2\boldsymbol{U}_{20}\Box\boldsymbol{B}_1\boldsymbol{U}_{10}\right]\label{MX:2D_X}\\\boldsymbol{Z} & = \boldsymbol{B}\boldsymbol{T}_{+} = \left[\boldsymbol{B}_2\boldsymbol{U}_{20}\Box\boldsymbol{B}_1\boldsymbol{U}_{1+}\mid\boldsymbol{B}_2\boldsymbol{U}_{2+}\Box\boldsymbol{B}_1\boldsymbol{U}_{10}\mid\boldsymbol{B}_2\boldsymbol{U}_{2+}\Box\boldsymbol{B}_1\boldsymbol{U}_{1+}\right]\label{MX:2D_Z}.
\end{align}
Moreover, $\boldsymbol{\theta} = \boldsymbol{T}\left(\boldsymbol{\beta}^{\top}, \boldsymbol{\alpha}^{\top}\right)^{\top} = \boldsymbol{T}_{0} \boldsymbol{\beta} + \boldsymbol{T}_{+}\boldsymbol{\alpha}$. We now obtain the adaptive penalty (and thus the precision matrix) associated with the new vector of coefficients. First, we express the full two-dimensional adaptive penalty (\ref{MX:add_penalty_2d_reduced}) in terms of $\left(\boldsymbol{\beta}^{\top}, \boldsymbol{\alpha}^{\top}\right)$ 

\begin{align*}
\left(\boldsymbol{\beta}^{\top},\boldsymbol{\alpha}^{\top}\right)\Biggl(&\sum_{u = 1}^{p_{11}p_{12}}\xi_{u}\boldsymbol{T}^{\top}\left(\mathbf{I}_{d_2}\otimes\boldsymbol{D}_{q_1}\right)^{\top}\mbox{diag}\left(\boldsymbol{\psi}_{u}\right)\left(\mathbf{I}_{d_2}\otimes\boldsymbol{D}_{q_1}\right)\boldsymbol{T} \\
&\sum_{s = 1}^{p_{21}p_{22}}\widetilde{\xi}_{s}\boldsymbol{T}^{\top}\left(\boldsymbol{D}_{q_2}\otimes\mathbf{I}_{d_1}\right)^{\top}\mbox{diag}\left(\widetilde{\boldsymbol{\psi}}_{s}\right)\left(\boldsymbol{D}_{q_2}\otimes\mathbf{I}_{d_1}\right)\boldsymbol{T}\Biggr)\left(\boldsymbol{\beta}^{\top}, \boldsymbol{\alpha}^{\top}\right)^{\top}.
\end{align*}
With some effort, it can be shown that the previous expression reduces to
\begin{align*}
\boldsymbol{\alpha}^{\top}\Biggl(&\sum_{u = 1}^{p_{11}p_{12}}\xi_{u}\boldsymbol{T}_{+}^{\top}\left(\mathbf{I}_{d_2}\otimes\boldsymbol{D}_{q_1}\right)^{\top}\mbox{diag}\left(\boldsymbol{\psi}_{u}\right)\left(\mathbf{I}_{d_2}\otimes\boldsymbol{D}_{q_1}\right)\boldsymbol{T}_{+} \\
&\sum_{s = 1}^{p_{21}p_{22}}\widetilde{\xi}_{s}\boldsymbol{T}_{+}^{\top}\left(\boldsymbol{D}_{q_2}\otimes\mathbf{I}_{d_1}\right)^{\top}\mbox{diag}\left(\widetilde{\boldsymbol{\psi}}_{s}\right)\left(\boldsymbol{D}_{q_2}\otimes\mathbf{I}_{d_1}\right)\boldsymbol{T}_{+}\Biggr)\boldsymbol{\alpha}.
\end{align*}
%That is to say, only vector $\boldsymbol{\alpha}$ in (\ref{MX:2DMix}) is penalised, while vector $\boldsymbol{\beta}$ does not contribute to the penalty (it is not penalised). In the equivalent mixed model, it implies that $\boldsymbol{\beta}$ is a vector of fixed effects, and that $\boldsymbol{\alpha}$ is a vector of random effects, assumed to be distributed according to a multivariate Gaussian with zero mean and precision matrix (the inverse of the variance-covariance matrix) given by
That is to say, only vector $\boldsymbol{\alpha}$ in (\ref{MX:2DMix}) contributes to the penalty, i.e., it is penalised, while $\boldsymbol{\beta}$ it is not. In the equivalent mixed model, it implies that $\boldsymbol{\beta}$ is a vector of fixed effects, and that $\boldsymbol{\alpha}$ is a vector of random effects, assumed to be distributed according to a multivariate Gaussian with zero mean and precision matrix (the inverse of the variance-covariance matrix) given by
\begin{equation}
\boldsymbol{G}^{-1} = \sum_{u=1}^{p_{11}p_{12}}\sigma_u^{-2}\boldsymbol{\mathcal{G}}_u + \sum_{s=1}^{p_{21}p_{22}}\widetilde{\sigma}_s^{-2}\widetilde{\boldsymbol{\mathcal{G}}}_s,
\label{G_inv_2d} 
\end{equation}
where $\sigma_u^{2} = \phi/\xi_{u}$, $\widetilde{\sigma}_s^2 = \phi/\widetilde{\xi}_{s}$ (recall that $\phi$ is the dispersion parameter) and 
\begin{align*}
\boldsymbol{\mathcal{G}}_u & = \boldsymbol{T}_{+}^{\top}\left(\mathbf{I}_{d_2}\otimes\boldsymbol{D}_{q_1}\right)^{\top}\mbox{diag}\left(\boldsymbol{\psi}_{u}\right)\left(\mathbf{I}_{d_2}\otimes\boldsymbol{D}_{q_1}\right)\boldsymbol{T}_{+},\\
\widetilde{\boldsymbol{\mathcal{G}}}_s & = \boldsymbol{T}_{+}^{\top}\left(\boldsymbol{D}_{q_2}\otimes\mathbf{I}_{d_1}\right)^{\top}\mbox{diag}\left(\widetilde{\boldsymbol{\psi}}_{s}\right)\left(\boldsymbol{D}_{q_2}\otimes\mathbf{I}_{d_1}\right)\boldsymbol{T}_{+}.
\end{align*}
We note in passing that the dimension of $\boldsymbol{\mathcal{G}}_u$ ($u = 1, \ldots, p_{11}p_{12}$) and $\widetilde{\boldsymbol{\mathcal{G}}}_s$ ($s = 1, \ldots, p_{21}p_{22}$) is $(d_1 - q_1)(d_2 -q_2)\times (d_1 - q_1)(d_2 -q_2)$. The fact that the precision matrix $\boldsymbol{G}^{-1}$ in (\ref{G_inv_2d}) is linear in the precision parameters $\sigma_u^{-2}$ and $\widetilde{\sigma}_s^{-2}$, allows resorting to the recently proposed SOP method \citep{MXRA19} for estimation. SOP is based on applying the method of successive approximations to easy-to-compute estimate updates of the variance parameters. This feature makes the method very computationally efficient, which is essential in our setting: The number of variance parameters (or equivalently smoothing parameters) to be estimated may be very large (recall that, in the full adaptive penalty, it equals to $p_{11}p_{12} +  p_{21}p_{22}$, where $p_{mw}$ ($m,w = 1,2$) relate to the dimension of the marginal B-splines bases involved in (\ref{MX:smooth_ad_2d_1}) and (\ref{MX:smooth_ad_2d_2})). Another computationally demanding step in the estimation is the calculation of matrices $\boldsymbol{\mathcal{G}}_u$ and $\widetilde{\boldsymbol{\mathcal{G}}}_s$ (this step needs to be performed only once). Here, the procedure can be sped up by exploiting the Kronecker structure of matrix $\boldsymbol{T}_{+}$ through the use of Generalised Linear Array Methods \citep[GLAM,][]{Currie2006}. All in all, the computational cost associated with the estimation of the adaptive P-spline model will mostly be driven by the number of variance parameters ($p_{11}p_{12} +  p_{21}p_{22}$) but also by the dimensions of the marginal B-spline bases involved in \eqref{MX:PsplineM_2D_v2} (i.e., $d_1$ and $d_2$). More precisely, $d_1$ and $d_2$ determine not only the dimension of the system of equations to be solved, but also the dimension of matrices $\boldsymbol{\mathcal{G}}_u$ and $\widetilde{\boldsymbol{\mathcal{G}}}_s$. These matrices are involved in the estimation of the variance parameters \citep[see][for more details]{MXRA19}. For the adaptive penalty in one dimension the ``equivalent'' matrices are diagonal, which allows reducing even further the computational cost \citep[see][Sections 3.3 and 4.1]{MXRA19}. Unfortunately, this is not the case here. Summarising, in the multidimensional adaptive setting one should be aware that the computational cost associated with model's estimation heavily depends not only on the number of the variance parameters (i.e., on the dimensions of the marginal B-spline bases used to ``smooth'' the smoothing parameters) but also on the dimensions of the marginal B-spline bases used to represent the bivariate smooth surface (which impact both the dimension of the system of equations to be solved and the dimension of $\boldsymbol{\mathcal{G}}_u$ and $\widetilde{\boldsymbol{\mathcal{G}}}_s$). Finally, our experience also suggests that the convergence of the SOP method in the adaptive case is slower than in the standard anisotropic situation, also impacting the computing time.  %Thus, in the multidimensional adaptive setting the computational cost associated with the estimation of the variance parameters heavily depend not only on their number but also on the dimension of $\boldsymbol{\mathcal{G}}_u$ and $\widetilde{\boldsymbol{\mathcal{G}}}_s$ (and thus on the number of marginal B-spline bases).     
\section{Simulation Study}\label{simulation}
This section reports the results of a simulation study conducted to study the empirical performance of the multidimensional adaptive penalties described in Section \ref{multi_penalty} above. For conciseness, the study concentrates on the two-dimensional case, and only the full adaptive penalty specification is evaluated (expression \eqref{MX:add_penalty_2d_reduced}). We compare the performance of our proposal with that described in \cite{Krivobokova08} and implemented in the \texttt{R}-package \texttt{AdaptFit} \citep{adapt_fit}. Also, a non-adaptive two-dimensional P-spline model (i.e., a model with a standard anisotropic penalty; see \eqref{MX:penalty_iso_2d}) is explored. As for the adaptive case, estimation here is based on the SOP method. Comparisons among the three approaches are performed in terms of the Mean Square Error (MSE) and the computing time. All computations are performed in (64-bit) \texttt{R} 4.0.2 \citep{R20}, and a 2.40GHz $\times$ 4 Intel$^\circledR$ Core$\texttrademark$ i7 processor computer with 15.6GiB of RAM and Ubuntu 16.04 LTS operating system.
\subsection{Scenarios and Setup}
Three different scenarios are considered in this study. Namely,
\begin{description}
\item[Scenario I.] This scenario is classical in the context of adaptive P-splines in two dimensions, and it has been considered, among others, by \cite{Lang04, Crainiceanu07} and \cite{Krivobokova08}. Covariates $x_{i1}$ and $x_{i2}$ are simulated independently from a uniform distribution on the interval $\left[0,1\right]$, and 
\begin{equation*}
g\left(\mu_i\right) = \eta_i  = f\left(x_{i1}, x_{i2}\right) = \exp\left(-15\left((x_{i1}/2.2 - 0.2)^2 + (x_{i2}/50)^2\right)\right).
\end{equation*}
The response data $y_i$ is then generated as
\[
y_i = \eta_i + \varepsilon_i\;\;\mbox{where}\;\;\varepsilon_i\sim N\left(0, s^2\right)\;\;\mbox{with}\;\;s = \frac{1}{4}\left(\min(f) - \max(f)\right).
\]
\item[Scenario II.] The second scenario corresponds to a situation where adaptive smoothing might be necessary. Covariates $x_{i1}$ and $x_{i2}$ are simulated independently from uniform distributions on the intervals $\left[-5,1.5\right]$ and $\left[-50,150\right]$, respectively, and  
\begin{equation*}
\eta_i  = f\left(x_{i1}, x_{i2}\right) = \exp\left(-15\left((x_{i1}/2.2 - 0.2)^2 + (x_{i2}/50)^2\right)\right).
\end{equation*}
Here, $y_i$ is generated under two different distributions
\begin{itemize}
	\item $y_i = \eta_i + \varepsilon_i$, where $\varepsilon_i\sim N\left(0, s^2\right)$ with $s \in \left\{0.1; 0.5\right\}$.
	\item $y_i \sim Bernoulli\left(p_i\right)$, with $p_i = \exp\left(\widetilde{\eta_i}\right)/\exp\left(1 + \widetilde{\eta_i}\right)$, where $\widetilde{\eta_i} = 6\eta_i - 3$,
\end{itemize}
	where the scaling factors that appear in the Bernoulli case are used to control the signal-to-noise ratio.
\item[Scenario III.] The last scenario aims to evaluate the possible loss of efficiency as a result of the use of a two-dimensional adaptive penalty when it is not needed. In this case, $x_{i1}$ and $x_{i2}$ are simulated independently from a uniform distribution on the interval $\left[0,1\right]$, and 
\begin{equation*}
\eta_i  = f\left(x_{i1}, x_{i2}\right) = 1.9(1.45 + \exp(x_{i1})\sin(13(x_{i1} - 0.6)^2))\exp(-x_{i2})\sin(7x_{i2}).
\end{equation*}
As for scenario II, $y_i$ is generated according to
\begin{itemize}
	\item $y_i = \eta_i + \varepsilon_i$, where $\varepsilon_i\sim N\left(0, s^2\right)$ with $s \in \left\{0.5; 2\right\}$.
	\item $y_i \sim Bernoulli\left(p_i\right)$, with $p_i = \exp\left(\eta_i\right)/\exp\left(1 + \eta_i\right)$.
\end{itemize}
\end{description}

\begin{figure}[h!]
 \begin{center}
 	\includegraphics[width=5.3cm]{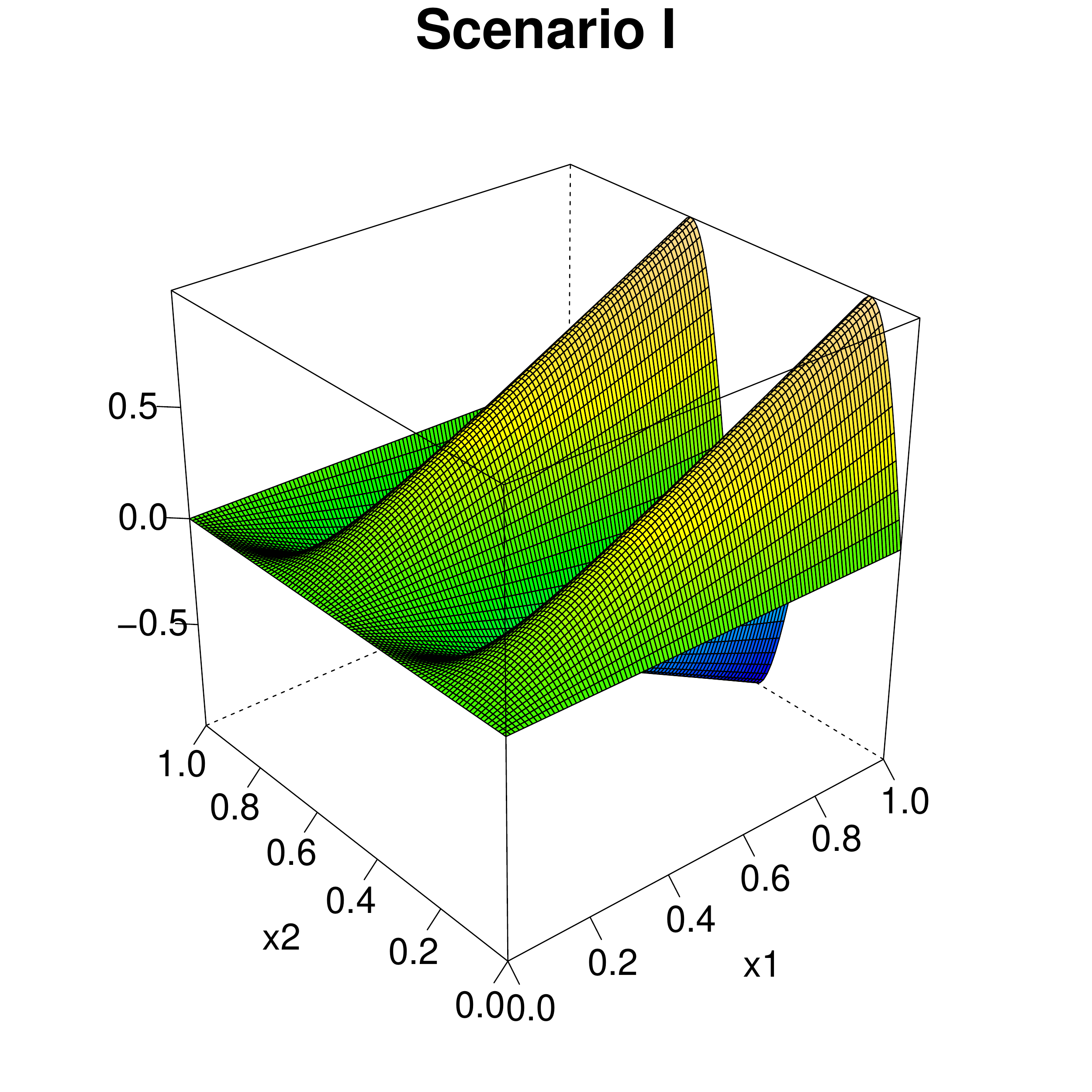}
  	\includegraphics[width=5.3cm]{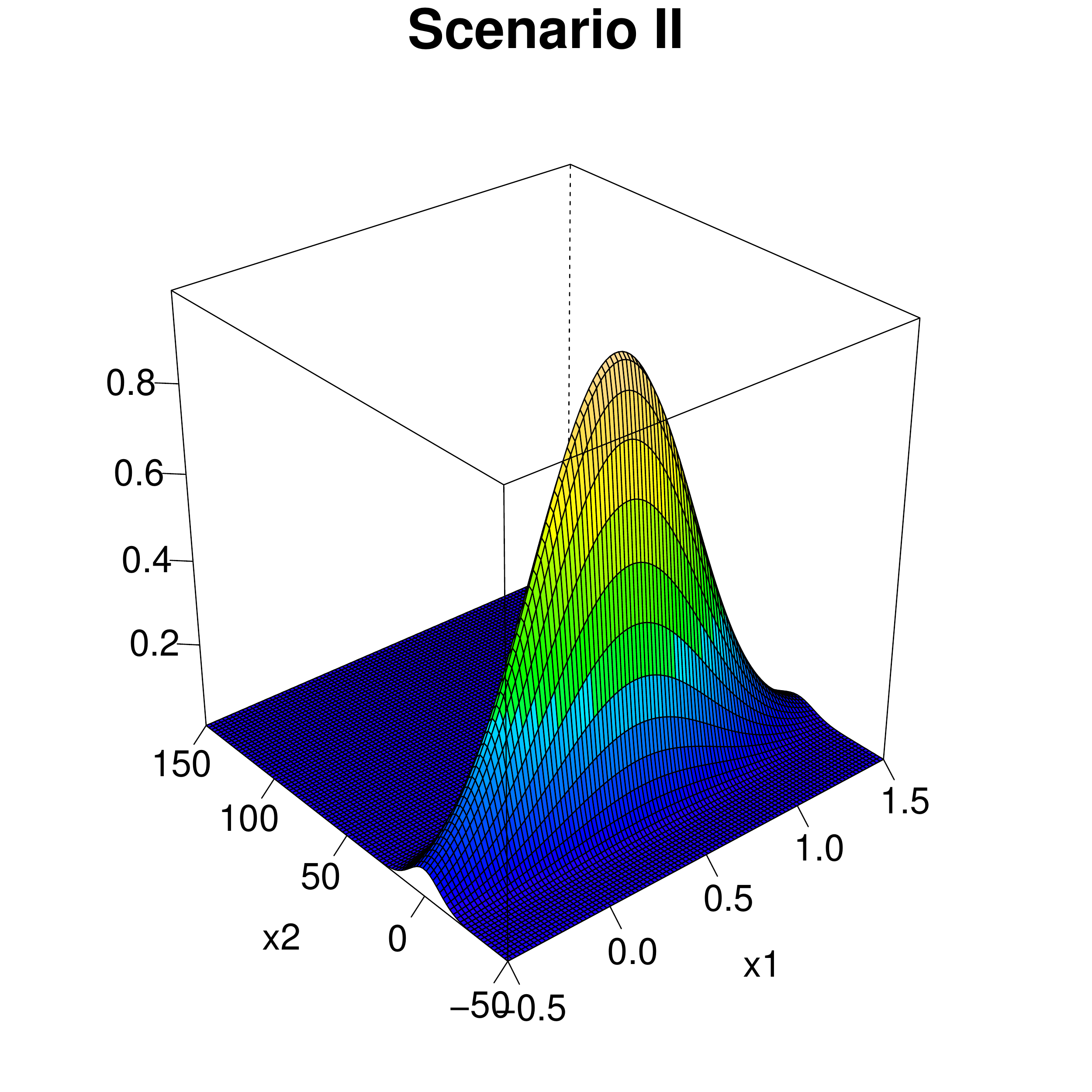}
   	\includegraphics[width=5.3cm]{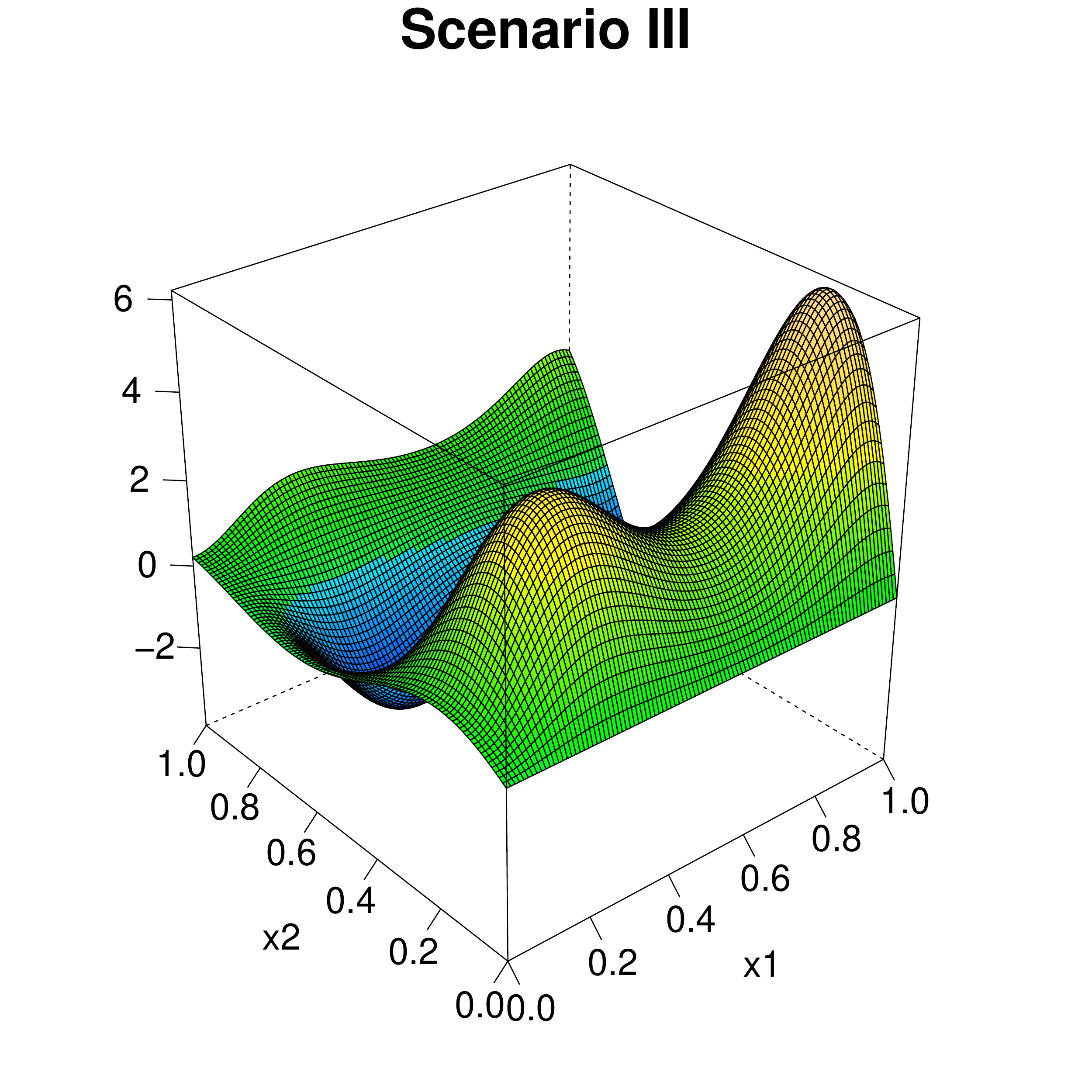}
	\end{center}
	\caption{For the simulation study: True two-dimensional functions used in Scenario I (left), II (middle) and III (right).}
	\label{true_functions}
\end{figure}

The true two-dimensional functions used in each scenario are shown in Figure \ref{true_functions}. For all scenarios, a total of $R = 250$ replicates are performed. Following previous studies, for Scenario I we only consider a sample size of $n = 300$, whereas, for Scenario II and III, results for $n = 1000$ are also studied. For the P-spline models with and without adaptive smoothing, we use second-order differences ($q_m = 2$) and marginal cubic B-splines bases of dimension $d_m = 12$ to represent the two-dimensional functions. For the adaptive approach, we consider the full adaptive penalty and choose $p_{mw} = 5$ ($m,w = 1,2$; see (\ref{MX:smooth_ad_2d_1}) and (\ref{MX:smooth_ad_2d_2})). This gives rise to a total of $50$ ($2 \times 5^2$) smoothing parameters (or variance components). For the proposal by \cite{Krivobokova08} (hereafter denoted as AdaptFit), we use the same specification as in that paper, i.e., $144$ ($12 \times 12$) knots for the two-dimensional function and $25$ ($5 \times 5$) knots for the variance parameters. Here, low-rank radial basis functions are used, and the knots are selected based on the \textit{clara} algorithm by \cite{KaufmanR90}. The evaluation of the practical performance of all approaches is judged by the MSE, computed at the observed covariate values. For Gaussian data, the true linear predictor ($\eta_i$) is chosen as the target. In the case of binary data, the MSE is computed on the response scale (the probability).
\subsection{Results}
\begin{figure}[h!]
 \begin{center}
	\includegraphics[width=6cm]{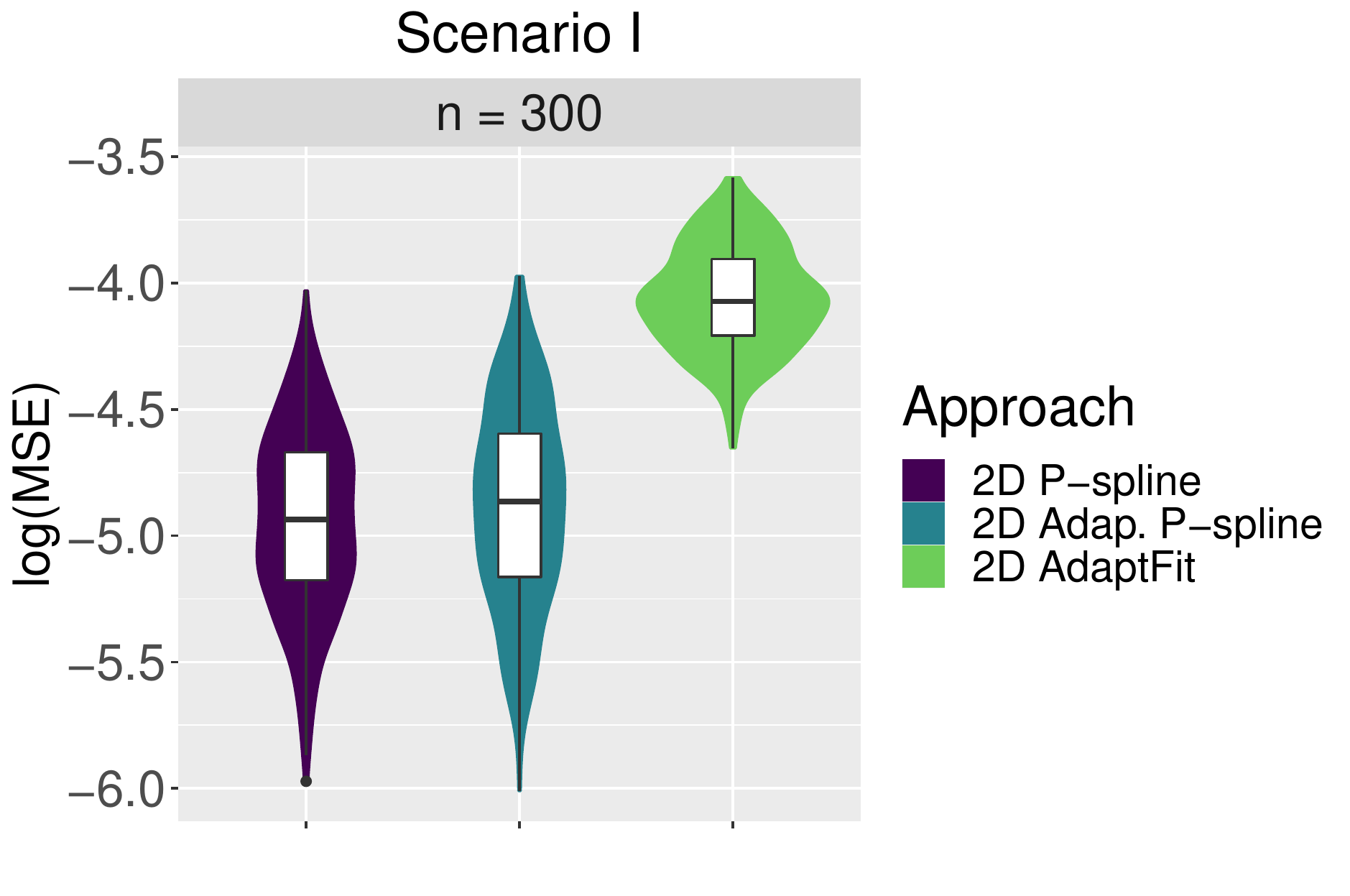}
 	\includegraphics[width=16cm]{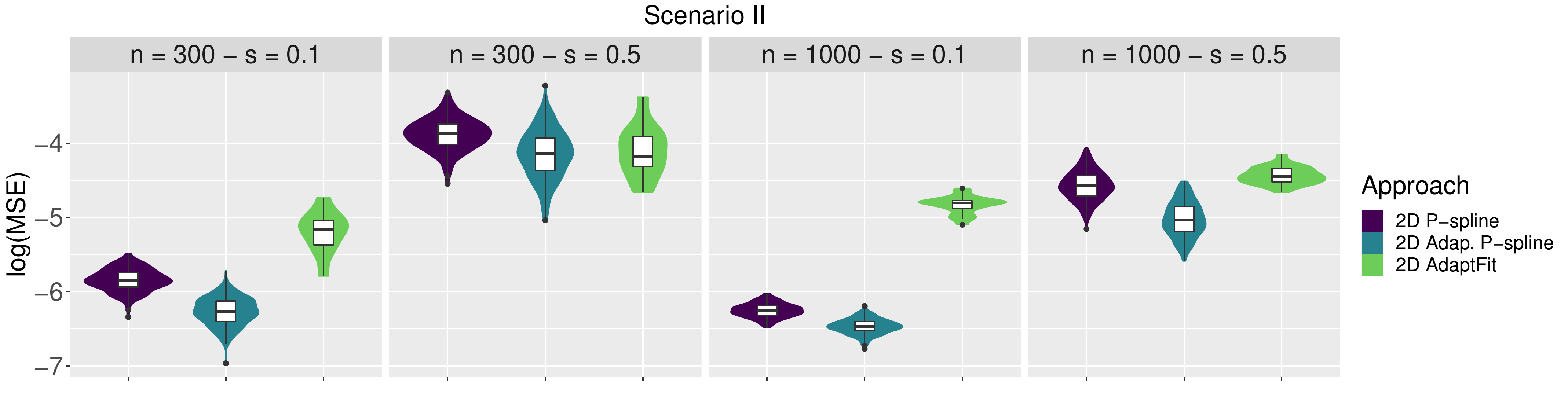}
  	\includegraphics[width=16cm]{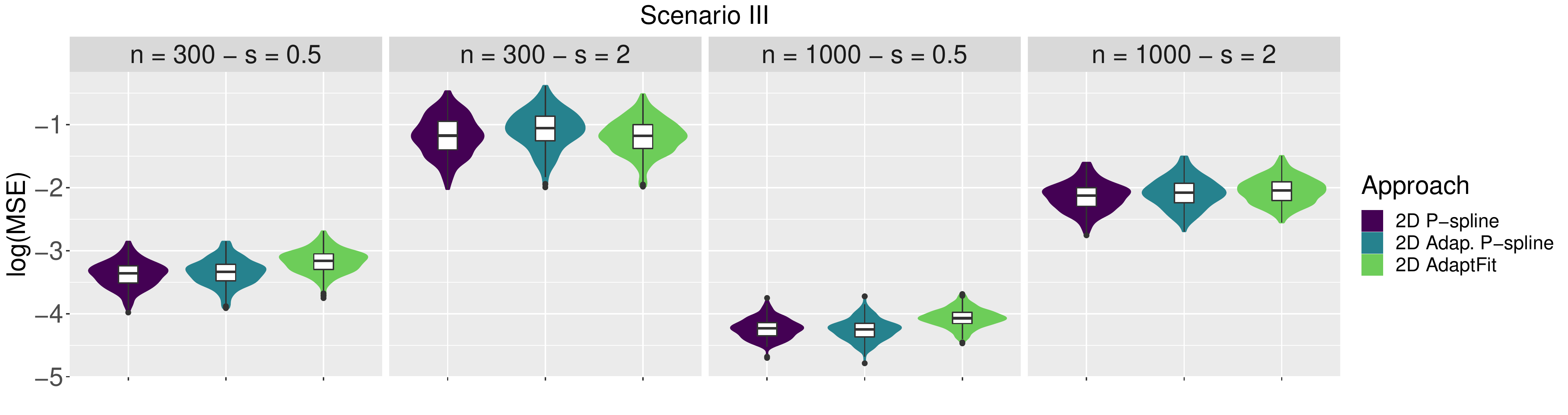}
	\end{center}
	\caption{For the simulation study: Boxplots of log(MSE) across $R = 500$ replicates for Scenarios I, II and III, Gaussian distribution, different levels of noise ($s$) and sample sizes ($n$). ``2D P-spline'' stands for the two-dimensional P-spline model with the standard anisotropic penalty, ``2D Adapt. P-spline'' for the model considering the full adaptive penalty in two-dimensions proposed in this paper, and ``2D AdaptFit'' for the proposal by \cite{Krivobokova08}. For ``2D AdaptFit'' results are based on those runs where the method converged (see main text).}
	\label{scenario_results_MSE_Gaussian}
\end{figure}

\begin{figure}[h!]
 \begin{center}
	\includegraphics[width=6cm]{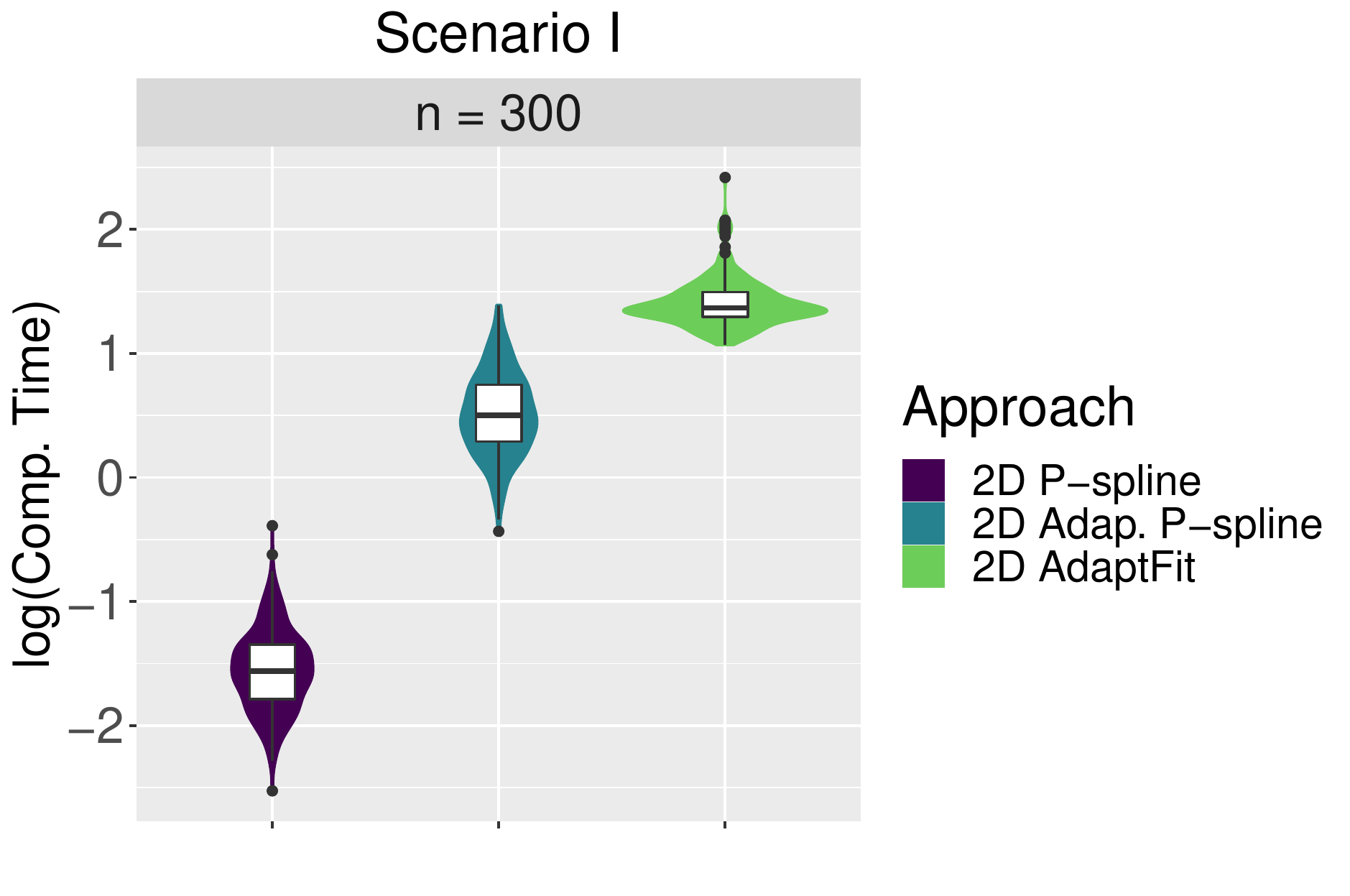}
 	\includegraphics[width=16cm]{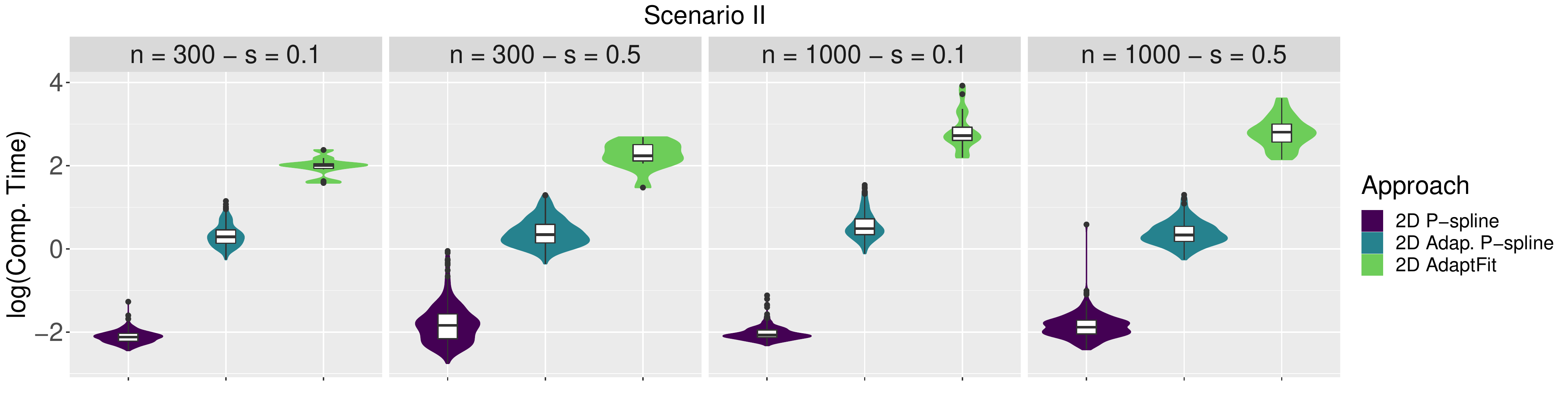}
  	\includegraphics[width=16cm]{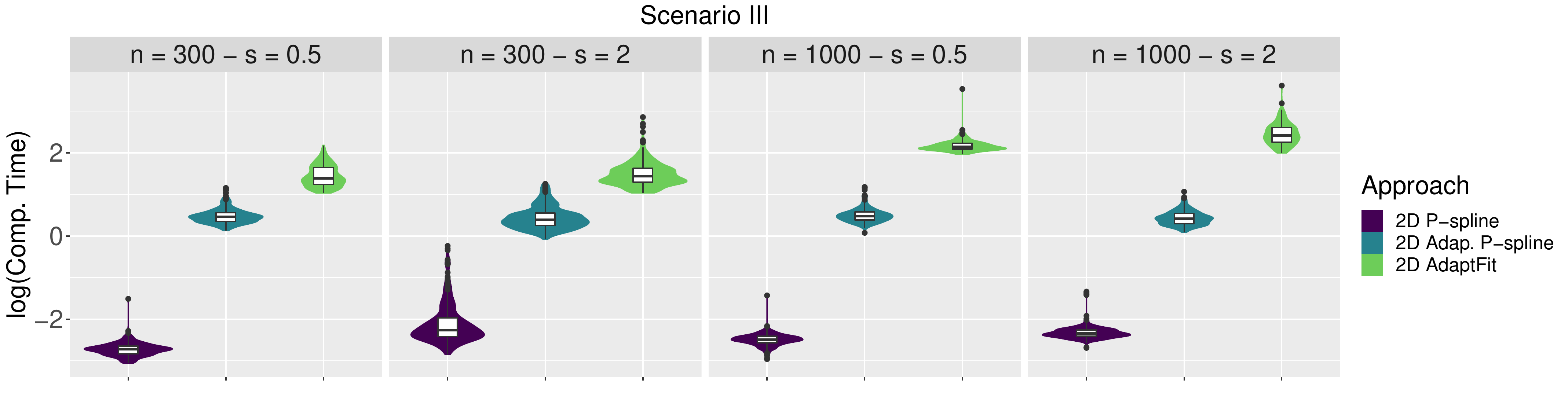}
	\end{center}
	\caption{For the simulation study: Boxplots of log(computing times) (in seconds) across $R = 250$ replicates for Scenarios I, II and III, Gaussian distribution, different levels of noise ($s$) and sample sizes ($n$). ``2D P-spline'' stands for the two-dimensional P-spline model with the standard anisotropic penalty, ``2D Adapt. P-spline'' for the model considering the full adaptive penalty in two-dimensions proposed in this paper, and ``2D AdaptFit'' for the proposal by \cite{Krivobokova08}. For ``2D AdaptFit'' results are based on those runs where the method converged (see main text).}
	\label{scenario_results_comp_time_Gaussian}
\end{figure}

Figure \ref{scenario_results_MSE_Gaussian} shows, for the three scenarios, boxplots of log(MSE) for the Gaussian distribution, and the different sample sizes and noise levels considered in the study. We start with the results for Scenario I. To our surprise, in this case, the best approach is the P-spline model with the standard anisotropic penalty, followed closely by the adaptive approach proposed in this paper. AdaptFit is the one performing the worst. This result somehow contradicts that reported in other papers \citep{Lang04}, where the adaptive approach (different to the one proposed here) performs better than the non-adaptive counterpart. In any case, we highlight that the results we obtain for AdaptFit are slightly better than those presented in \cite{Krivobokova08} (median of log(MSE) of $-4.07$ and $-3.79$, respectively), which in turn outperform results in both \cite{Lang04} and \cite{Crainiceanu07}. If we focus on the results for Scenario II (the one requiring, in principle, adaptive smoothing), we see that the full two-dimensional adaptive approach proposed in this paper is the one that performs the best, for all sample sizes and noise levels. We should note that, for this scenario, AdaptFit presents severe convergence problems, notwithstanding we increase to $1000$ the maximum number of iterations for the estimation of both the mean function and the variance of the random effects. The number of runs for which the method converges ranges from $47$ (out of $250$ for $n = 1000$ and $s = 0.5$) to $10$ ($n = 300$ and $s = 0.1$). Thus, for this approach, the results shown in Figure \ref{scenario_results_MSE_Gaussian} correspond to those runs where the method converges, and may, therefore, be misleading. Regarding Scenario III (which does not require adaptive smoothing), the three approaches perform more or less similarly, although the non-adaptive P-spline model presents slightly better performance. The message here is that the extra flexibility afforded by the adaptive approach does not translate into an important loss in efficiency. For this scenario, AdaptFit also shows convergence problems, but for a limited number of runs (between $2$ and $3$). In Figure \ref{scenario_results_comp_time_Gaussian}, results regarding computing times (in seconds) are presented (in logarithm scale). As could have been expected, in all cases, fitting the non-adaptive P-spline model is faster than using the adaptive alternatives, but our approach clearly outperforms AdaptFit. To give some numbers, for one of the worst-case situations for our adaptive approach (Scenario III, $n = 1000$ and $s = 2$), the standard anisotropic two-dimensional P-spline model requires, in median, around $0.1$ seconds, whereas our adaptive approach and the AdaptFit require around $1.5$ and $11.2$ seconds, respectively. In any case, computing times are more than reasonable. However, we are aware that these values cannot be considered as benchmarks: It is expected that computing times will increase significantly for other (and more ``generous'') bases specifications. For instance, when using $d_m = 16$ and $p_{mw} = 6$ (for AdaptFit we use $16 \times 16$ equally spaced knots for the two-dimensional function and $6 \times 6$ equally spaced knots for the variance parameters), the computing times increase to $0.5$, $6.5$ and $57.2$ (median), for, respectively, the standard anisotropic two-dimensional P-spline model, the proposed adaptive approach and AdaptFit. A more detailed study of the impact, in both MSE and computing time, of increasing $p_{mw}$ (while keeping constant $d_m$) can be found as online material. In brief, the results show that, as expected, an increase in $p_{mw}$ do have an impact in the computing time of our proposal; the larger $p_{mw}$, the more time consuming the estimation algorithm. For AdapFit (for those runs where it converges) the impact (of increasing the number of knots for the estimation of the variance components) is not as marked. Also, the results suggest that increasing $p_{mw}$ may lead to slightly worse results regarding MSE.

%Our simulations provide a median of log(MSE) of $-4.02$ while \cite{Krivobokova08} report a value of $-3.79$.

\begin{figure}[h!]
 \begin{center}
 	\includegraphics[width=16cm]{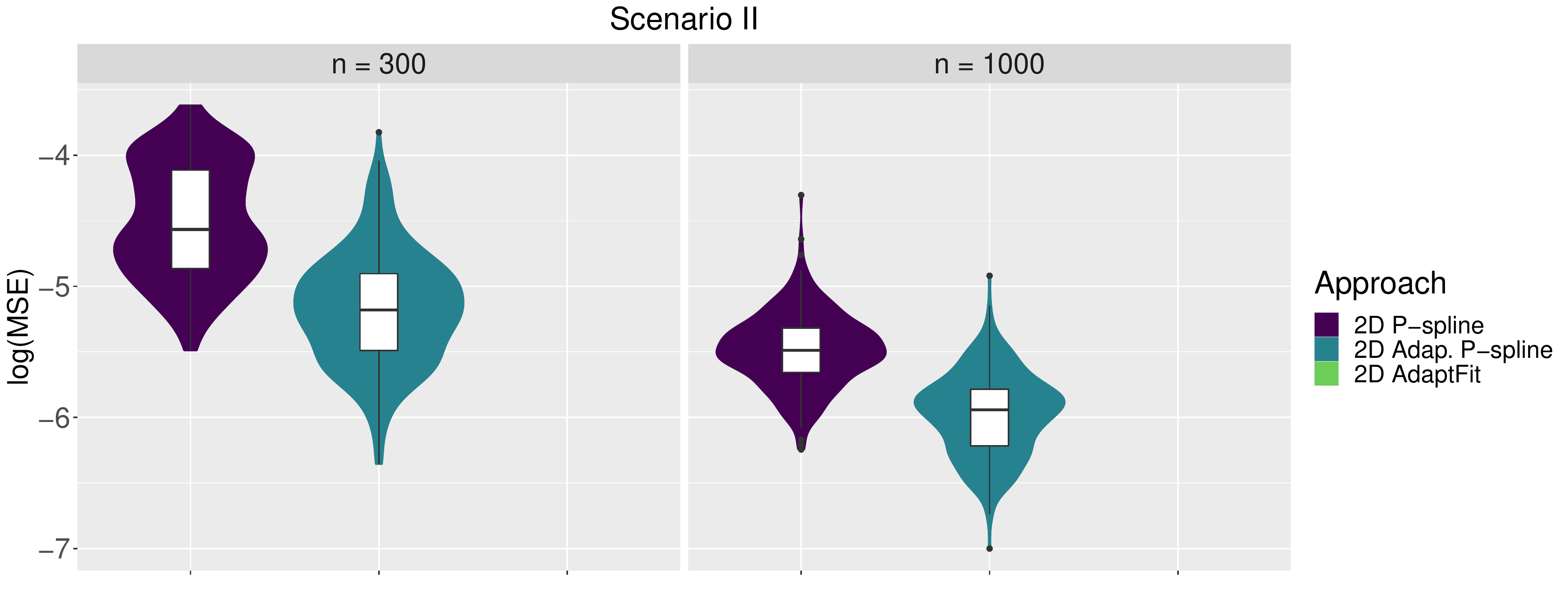}
  	\includegraphics[width=16cm]{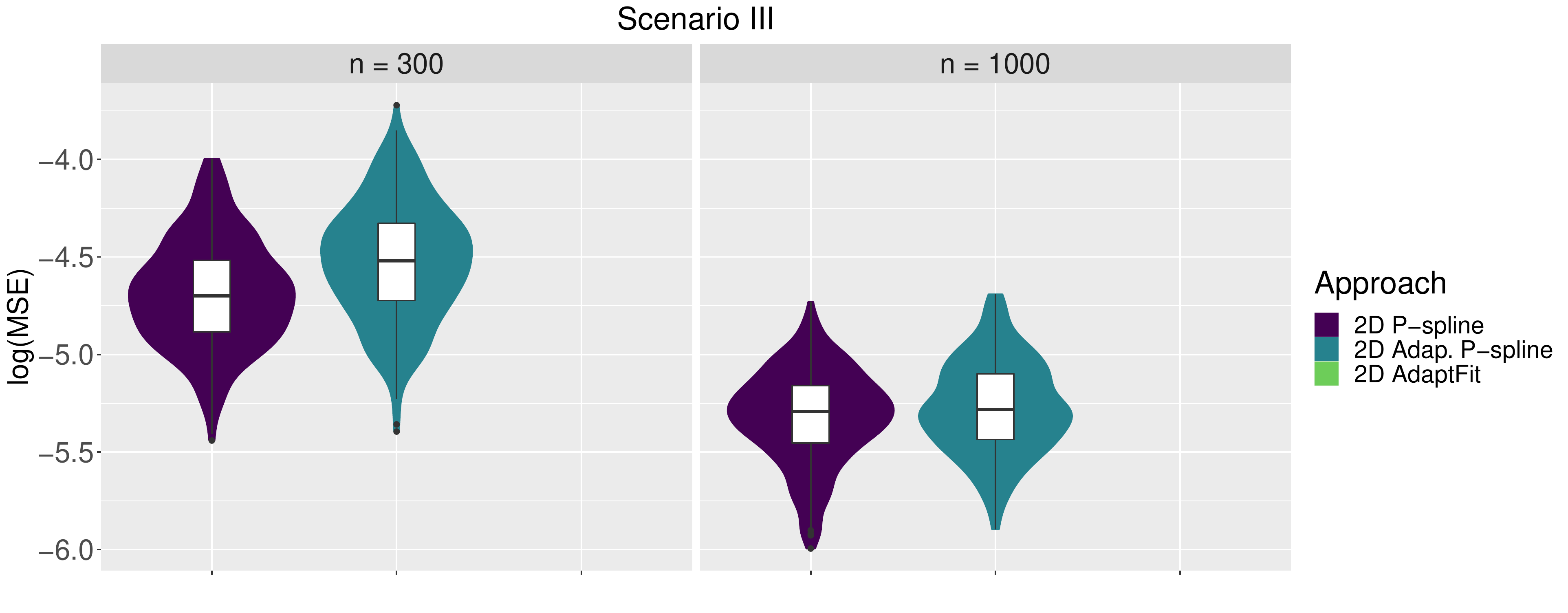}
	\end{center}
	\caption{For the simulation study: Boxplots of log(MSE) across $R = 250$ replicates for Scenarios II and III, Bernoulli distribution and different sample sizes ($n$). ``2D P-spline'' stands for the two-dimensional P-spline model with the standard anisotropic penalty, ``2D Adapt. P-spline'' for the model considering the full adaptive penalty in two-dimensions proposed in this paper, and ``2D AdaptFit'' for the proposal by \cite{Krivobokova08}. For 2D AdaptFit results are based on those runs where the method converged (see main text).}
	\label{scenario_results_MSE_bin}
\end{figure}

\begin{figure}[h!]
 \begin{center}
 	\includegraphics[width=16cm]{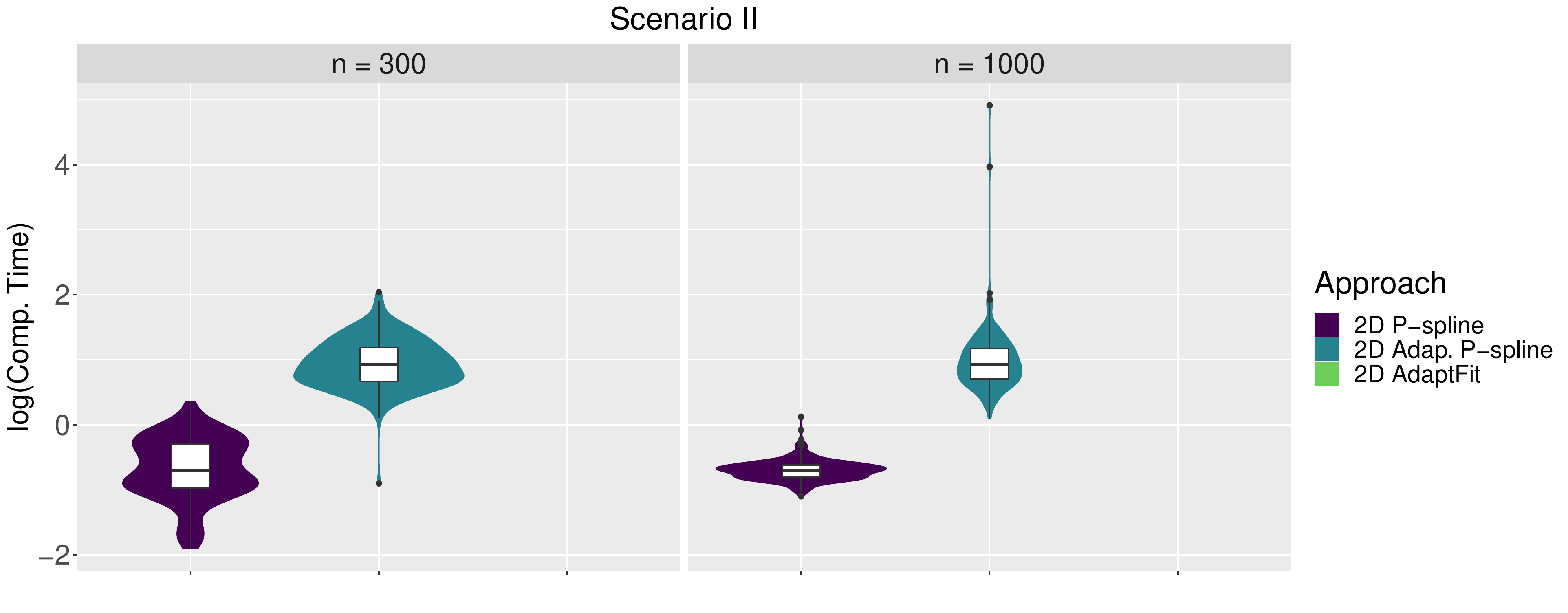}
  	\includegraphics[width=16cm]{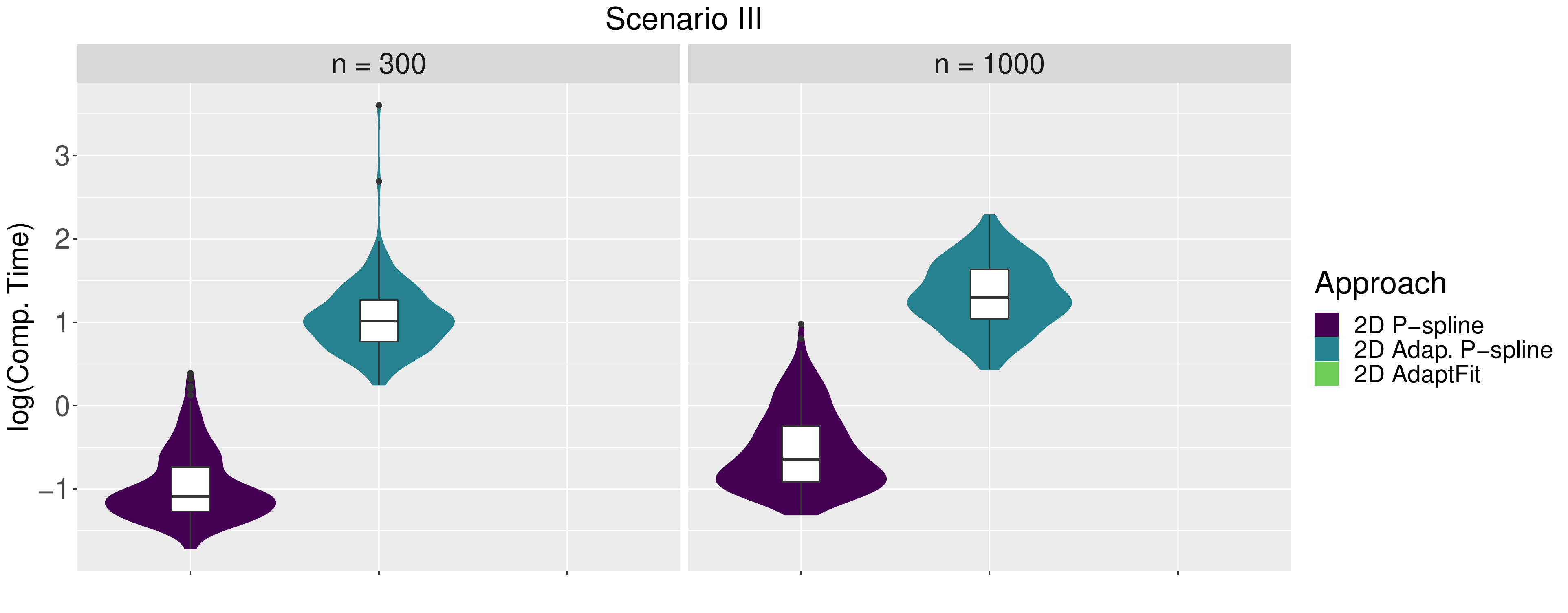}
	\end{center}
	\caption{For the simulation study: Boxplots of log(computing times) (in seconds) across $R = 250$ replicates for Scenarios II and III, Bernoulli distribution and different sample sizes ($n$). ``2D P-spline'' stands for the two-dimensional P-spline model with the standard anisotropic penalty, ``2D Adapt. P-spline'' for the model considering the full adaptive penalty in two-dimensions proposed in this paper, and ``2D AdaptFit'' for the proposal by \cite{Krivobokova08}. For ``2D AdaptFit'' results are based on those runs where the method converged (see main text).}
	\label{scenario_results_comp_time_bin}
\end{figure}

Results for Scenarios II and III and the Bernoulli distribution are shown in Figure \ref{scenario_results_MSE_bin} (log(MSE)) and Figure \ref{scenario_results_comp_time_bin} (log(computing times)). For both scenarios and sample sizes considered, the proposal by \cite{Krivobokova08} does not converge for any run. Again, for Scenario II the adaptive approach described in this paper outperforms (in terms of log(MSE)) the non-adaptive one, whereas for Scenario III and $n = 300$ it is the opposite. For $n = 1000$, both approaches behave similarly. This result seems to indicate that, for non-Gaussian data, the adaptive approach may require moderate to large sample sizes to behave similarly to the non-adaptive counterpart (when adaptive smoothing is not needed). Regarding computing times, the adaptive approach, is, as expected, more time-consuming, yet still very fast.  

\section{Applications}\label{Application}
In this section we show the potential of the multidimensional adaptive P-spline model proposed in this paper by applying it to two different datasets. One is the neurons' activity study described in detail in Section \ref{motivation}. Here, the data asks for a spatio-temporal analysis, and thus a three-dimensional adaptive P-spline model is considered. Before showing in detail the results for this study, we present an application in two dimensions. Namely, we apply our approach to the dataset on the absenteeism of workers discussed in \cite{Krivobokova08}. %that motivated the work and
   
\subsection{Absenteeism Data}\label{Absenteeism}
\begin{figure}[h!]
 \begin{center}
 	\includegraphics[width=6cm]{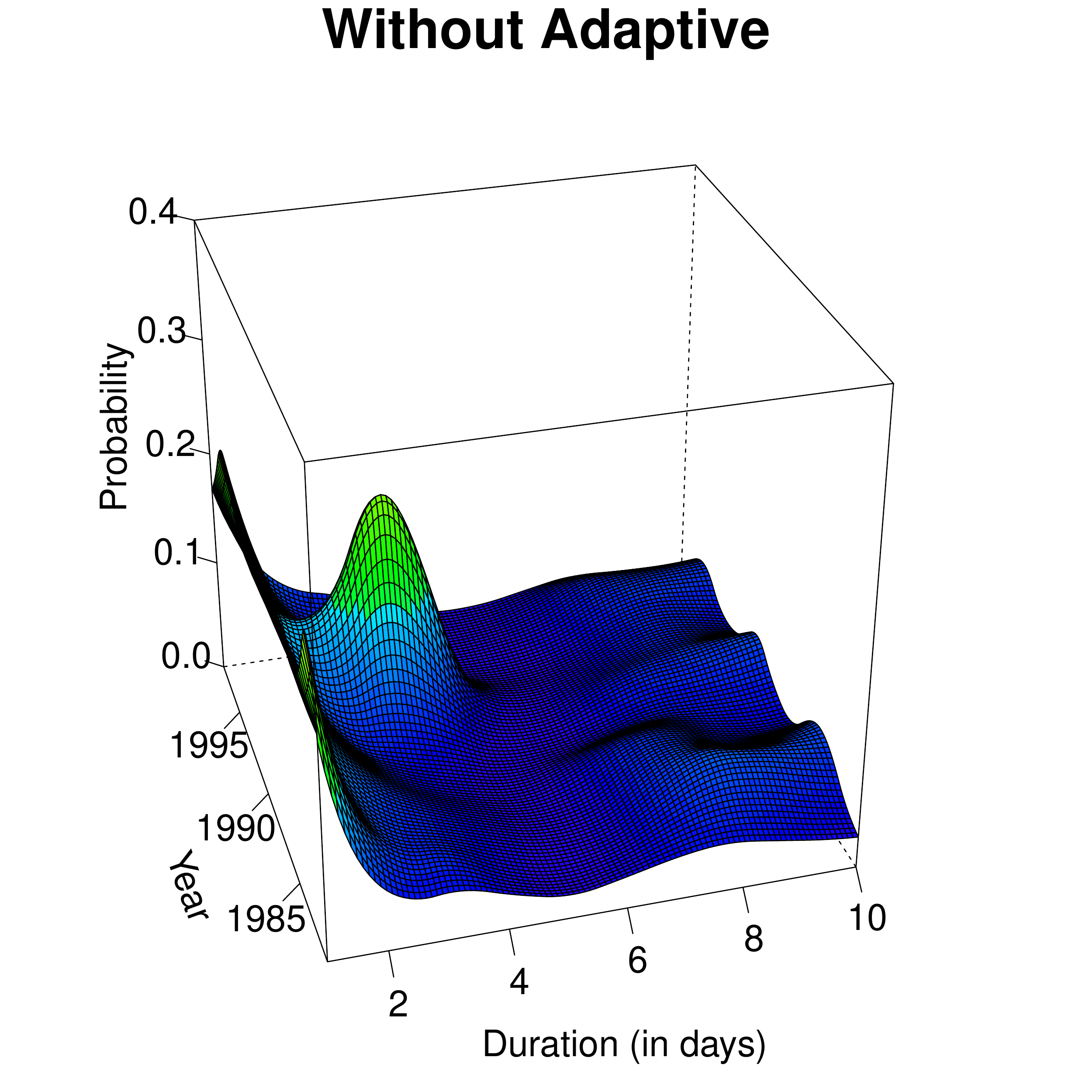}
  	\includegraphics[width=6cm]{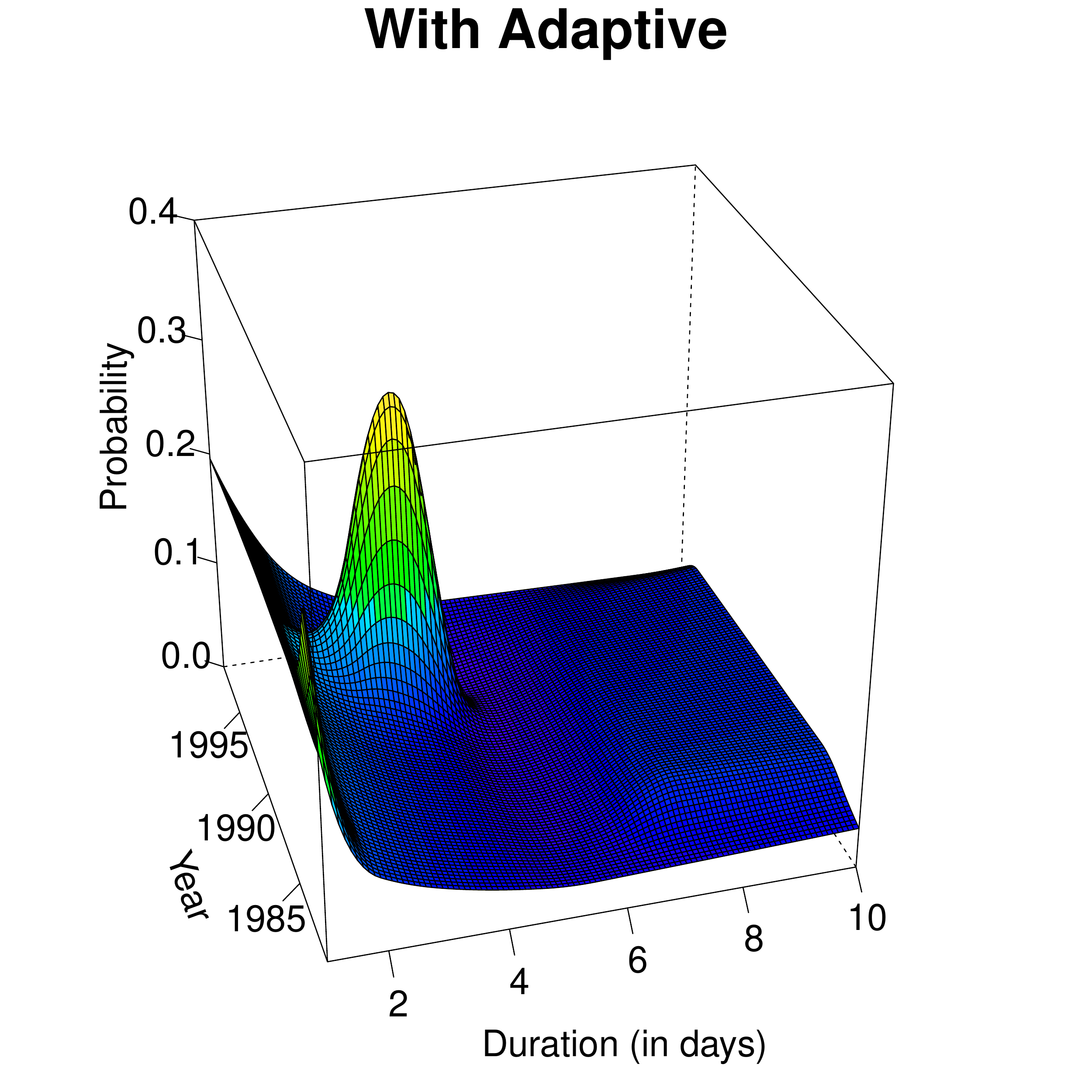}

	\vspace{7mm}
	\includegraphics[width=6cm]{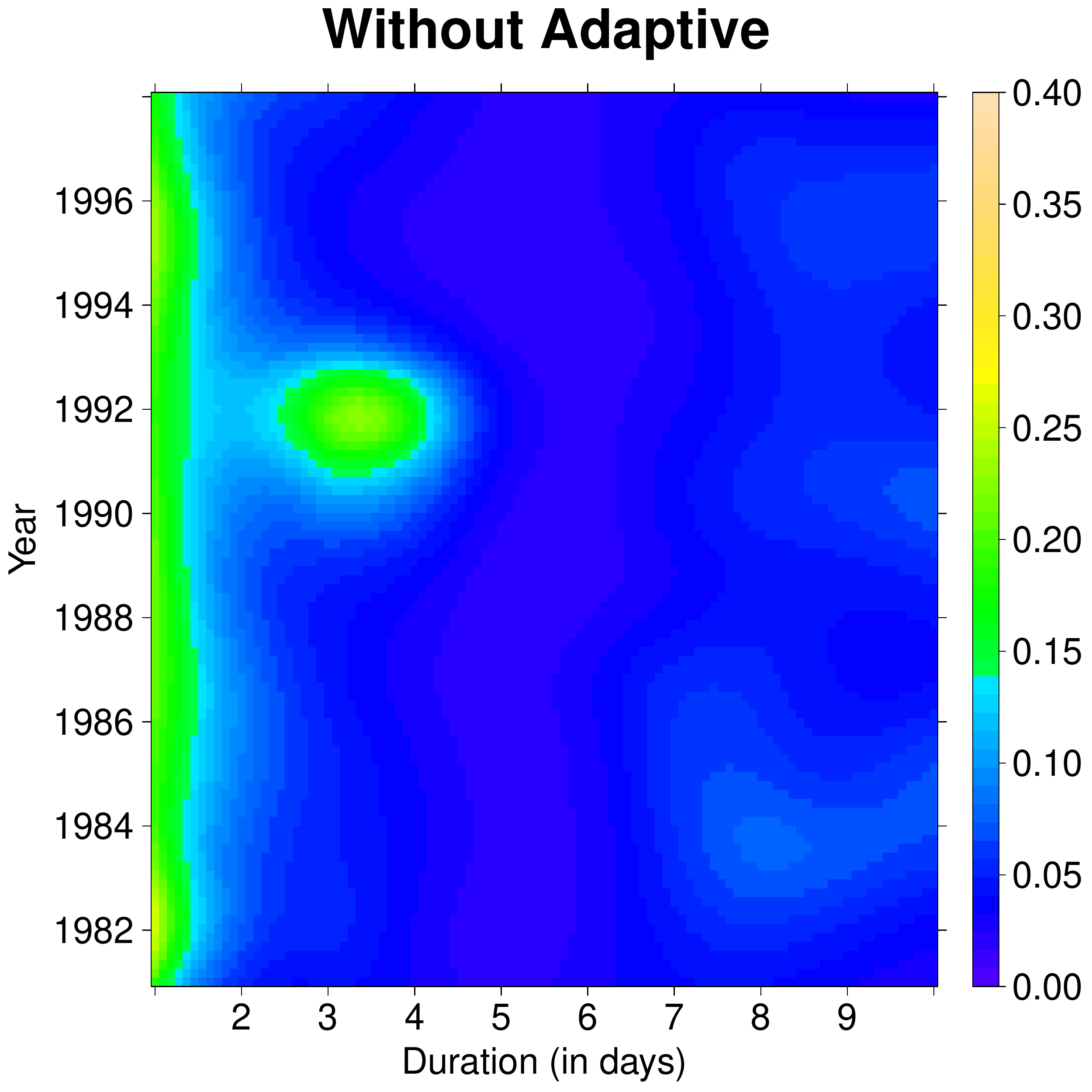}
 	\includegraphics[width=6cm]{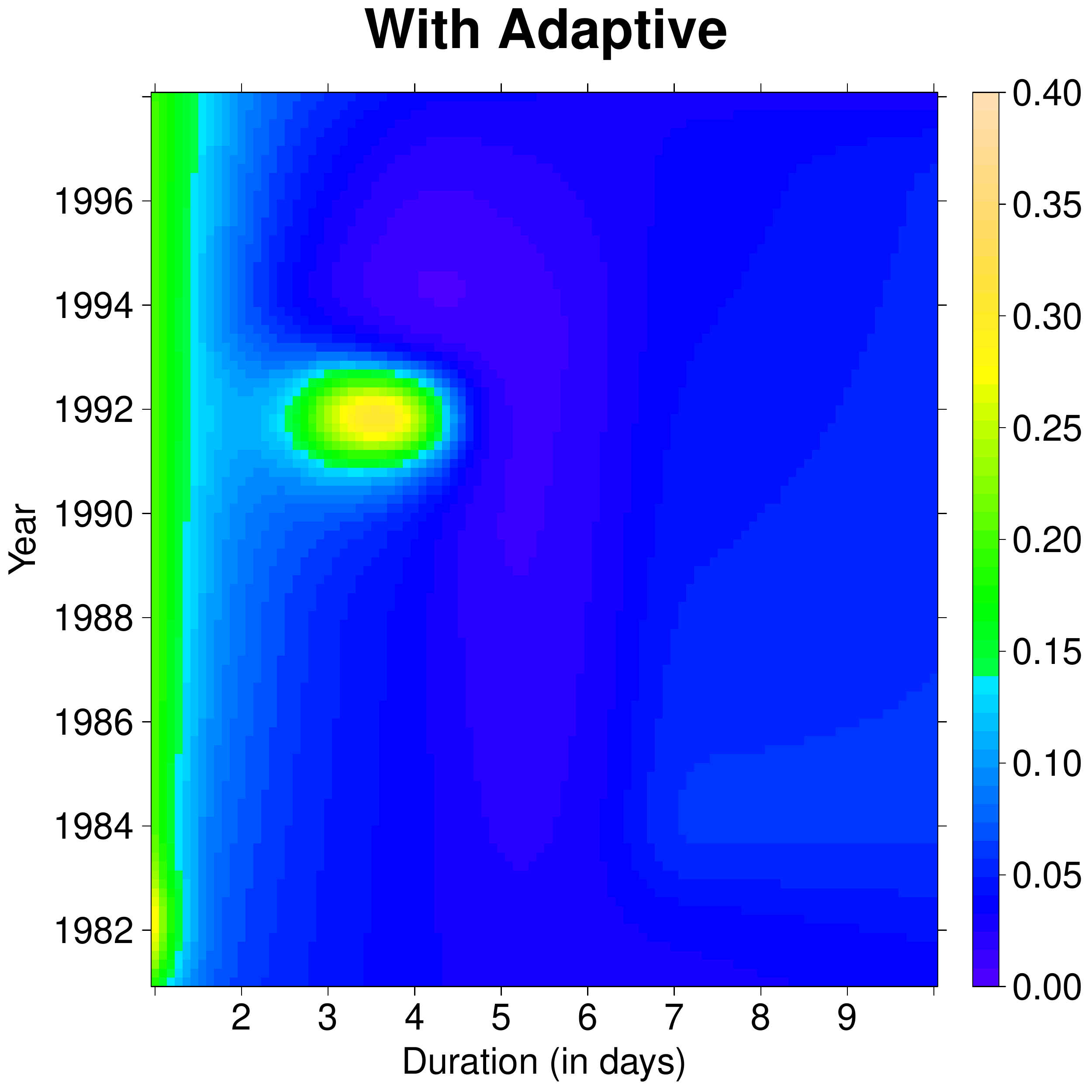}

	\end{center}
	\caption{For the absenteeism data: estimated $\Pr[\text{Duration} = \mathcal{D} \mid \text{Duration} \geq \mathcal{D}, \text{Year} = \mathcal{Y}]$. The left-hand side plots show the results using the standard (non-adaptive) anisotropic penalty, and the right-hand side plots those using the full two-dimensional adaptive penalty.}
	\label{abs_results}
\end{figure}

The dataset analysed here concerns absenteeism spells of workers at a company in Germany between $1981$ and $1998$. The interest lies in estimating the probability of returning to work after a sick leave of duration $\mathcal{D}$ and in studying if this probability depends on the year the sick leave took place, i.e.,
\[
\Pr[\text{Duration} = \mathcal{D} \mid \text{Duration} \geq \mathcal{D}, \text{Year} = \mathcal{Y}],
\]
with $\mathcal{D} \in \{1, \ldots, 10\}$ and  $\mathcal{Y} \in \{1981, \ldots, 1998\}$. Estimation of the above probability is based on a data augmentation strategy. The duration of each sick absence spell (i.e., the number of working days the worker has been absent), denoted as $d_i$, is ``expanded'' into a sequence of $d_i$ observations $y_{i1}, \ldots, y_{id_i}$, with $y_{i\mathcal{D}} = 0$ when $d_i < \mathcal{D}$ and $y_{id_i} = 1$. Two exceptions apply to the previous rule (censored observations): (1) if the duration of the sick leave is larger than $10$, the sequence is truncated at $10$, and thus $y_{i10} = 0$; and (2) if the last day of absenteeism and the first day of returning to work are not consecutive working days (e.g., there is a weekend in between), then $y_{id_i} = 0$ \cite[see][for more details]{Krivobokova08}. With this in mind, the model to be estimated is
\begin{equation}
\text{logit}\Pr[y_{i\mathcal{D}} = 1 \mid \mathcal{D}, yr_i] = f(\mathcal{D}, yr_i),
\label{model_abs}
\end{equation}
where $yr_i$ is the year when the $i$-th sick absence spell took place (it is the same value for the whole sequence of  observations $y_{i1}, \ldots, y_{id_i}$). The augmented dataset can be found in the \texttt{R}-package \texttt{AdaptFit} \citep{adapt_fit} as \texttt{absent}. For fitting the model, we consider the full two-dimensional adaptive penalty discussed in Section \ref{sec_2D}. Besides, we also fit a model with the standard anisotropic penalty (i.e., non-adaptive smoothing). In both cases, we use marginal cubic B-splines bases of dimension $d_m = 13$ to represent the smooth surface, jointly with second-order differences ($q_m = 2$). For the adaptive approach, we choose $p_{mw} = 10$ ($m,w = 1,2$), yielding a total of $200$ ($2 \times 10^2$) smoothing parameters (or variance components). These values are chosen to provide enough flexibility to the models. Figure \ref{abs_results} shows the results using both approaches. Regarding computing times, the full two-dimensional adaptive P-spline model needs $12$ seconds to be fitted, in contrast to $4$ seconds using the standard anisotropic penalty. Despite the increase, the computing time of the adaptive approach remains at a reasonable level. In terms of degrees of freedom or effective dimensions (ED), for the model without adaptive penalty, we obtain a total ED of $46.5$ (out of $200$), whereas with adaptive smoothing we have an ED of $16.7$ (out of $200$). We should note that using SOP the total ED is obtained by adding up the EDs associated with each smoothing/variance parameter in the model (plus the dimension of the fixed part). Details can be found in \cite{MXRA19}. For (multidimensional) adaptive P-spline models, most of the EDs (or equivalently the variance parameters) are almost zero, meaning a local linear fit (i.e., they ``allow'' the model to better adapt to the data regardless adaptivity is needed or not). Indeed, it is interesting to note that the flexibility afforded by the adaptive penalty does not translate into a more complex (and thus more prone to overfit) model. As indicated by the total EDs but also visible from the results shown in Figure \ref{abs_results}, it is the opposite. As can be seen, both approaches detect a peak around year $1992$ and an absenteeism duration of $3$ days, but the peak is more pronounced with the adaptive approach. More importantly, the adaptive approach avoids, for long absenteeism times, the wiggly (and possibly unrealistic) estimates obtained with the non-adaptive model. Our results are in concordance with those presented in \cite{Krivobokova08}. As discussed in that paper, the peak has a possible explanation. During $1992$ and $1993$, the company went through a downsizing process, and more than half of the employees were fired. The peak reflects that during the downsizing period the duration of sick leaves was shorter with more employees returning right after three days. The underlying reason might be that, according to the German law, workers who report sick for more than three consecutive working days have to provide a medical certificate. We finish this section by noting that we also fit model \eqref{model_abs} using the adaptive penalty simplifications described in Section \ref{simplifications}. We use the conditional Akaike Information Criterion (cAIC) proposed by \cite{DONOHUE11} to compare the fits provided by the different penalties. According to the cAIC, the best model is the one using the full two-dimensional adaptive penalty (results not shown). %We use the conditional Akaike Information Criterion \cite[cAIC,][]{Akaike1974,DONOHUE11} to compare the fits provided by the different penalties. According to the cAIC, the best model is the one using the full two-dimensional adaptive penalty (results not shown).

\subsection{Visual Receptive Fields}\label{app_vrc}
\begin{figure}[ht!]\centering
\includegraphics[width=15cm]{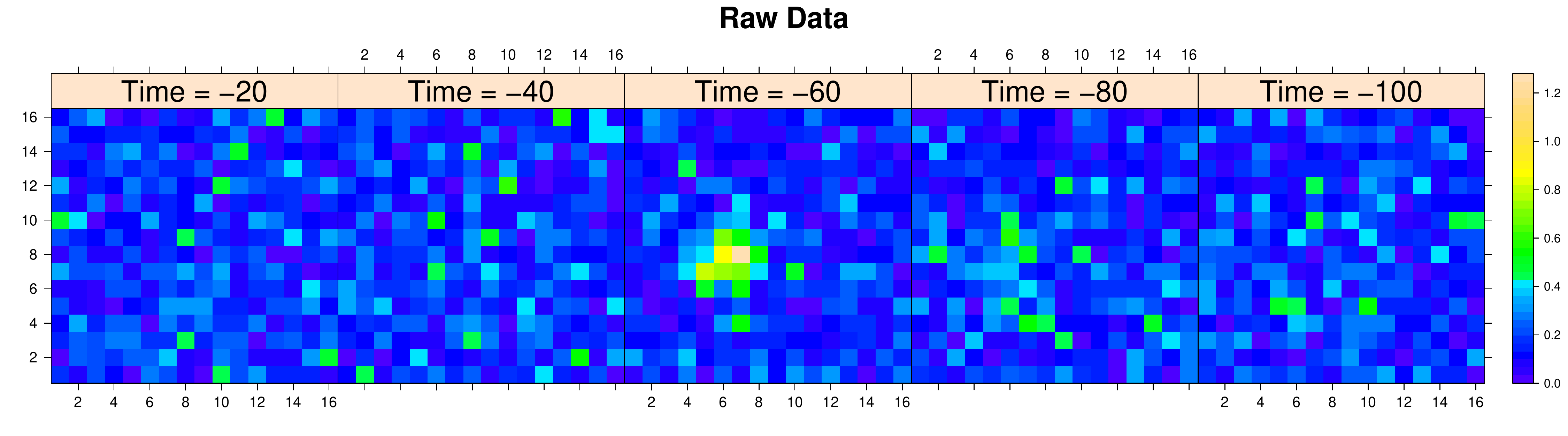}
\includegraphics[width=15cm]{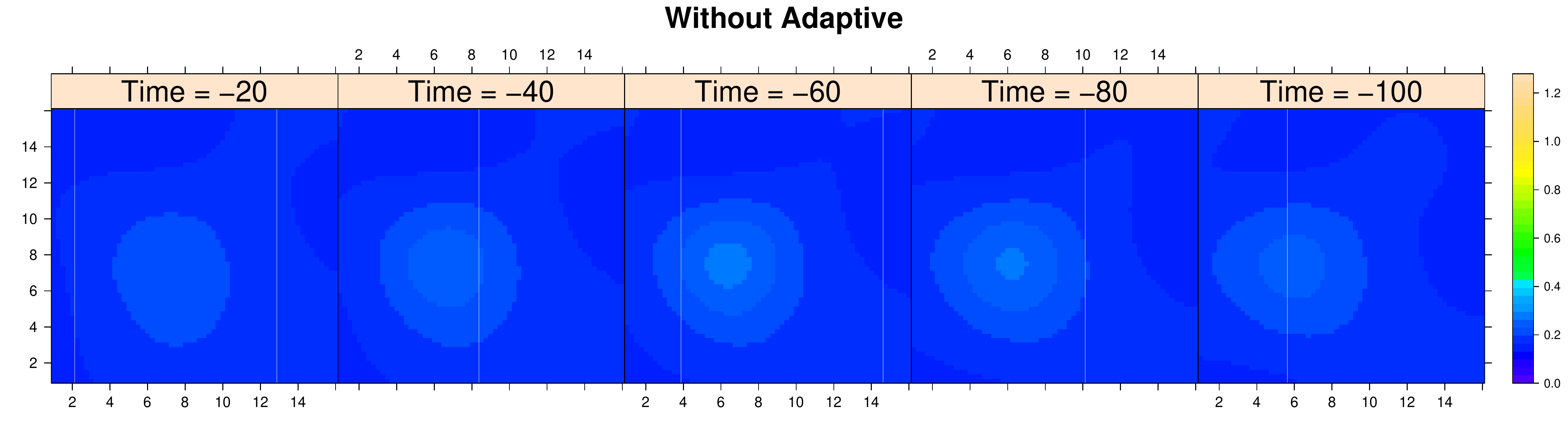}
\includegraphics[width=15cm]{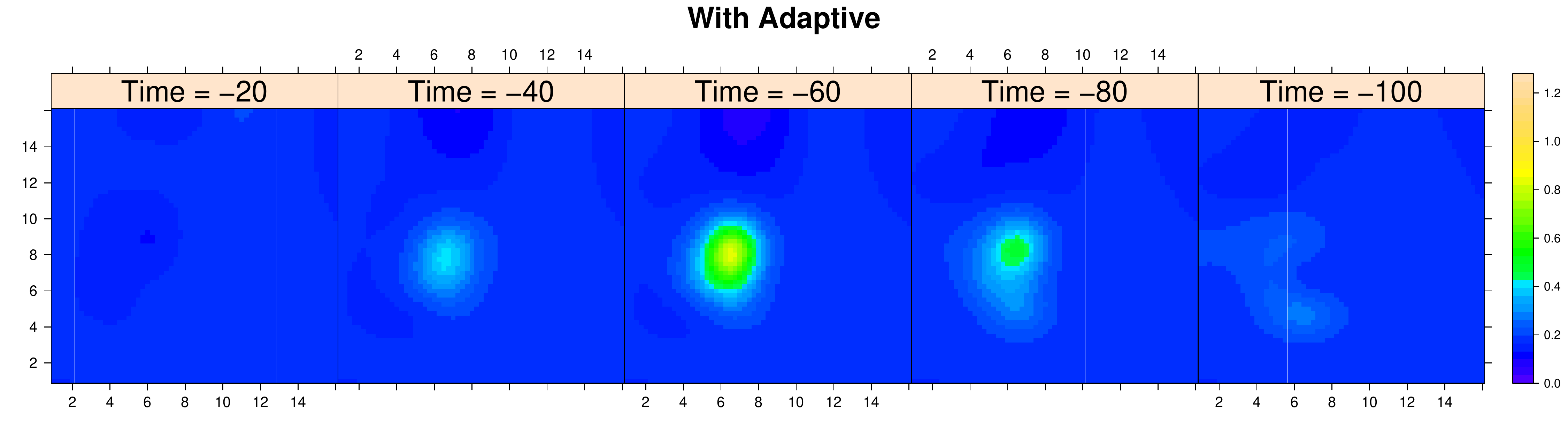}
\caption{For the visual receptive field study: Level plot of the the observed and smoothed ON-RFmaps (firing rates) for the right eye of cell FAU3. First row: observed. Second row: estimates without locally adaptive smoothing. Third row: estimates with locally adaptive smoothing. \label{MX:FAP1}}
\end{figure}

\begin{figure}[ht!]\centering
\includegraphics[width=15cm]{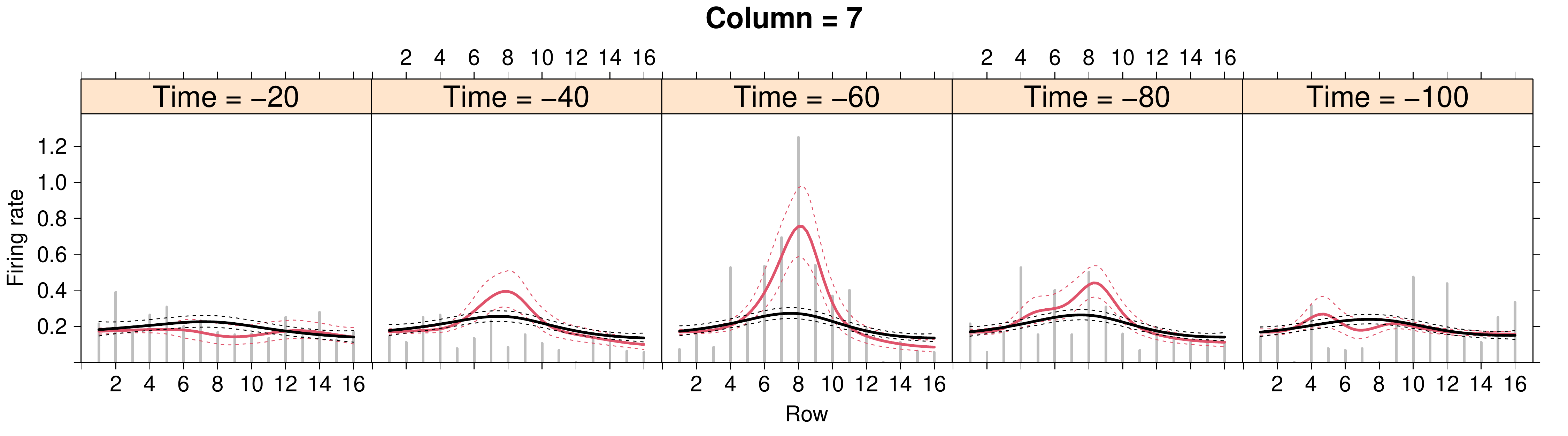}
\includegraphics[width=15cm]{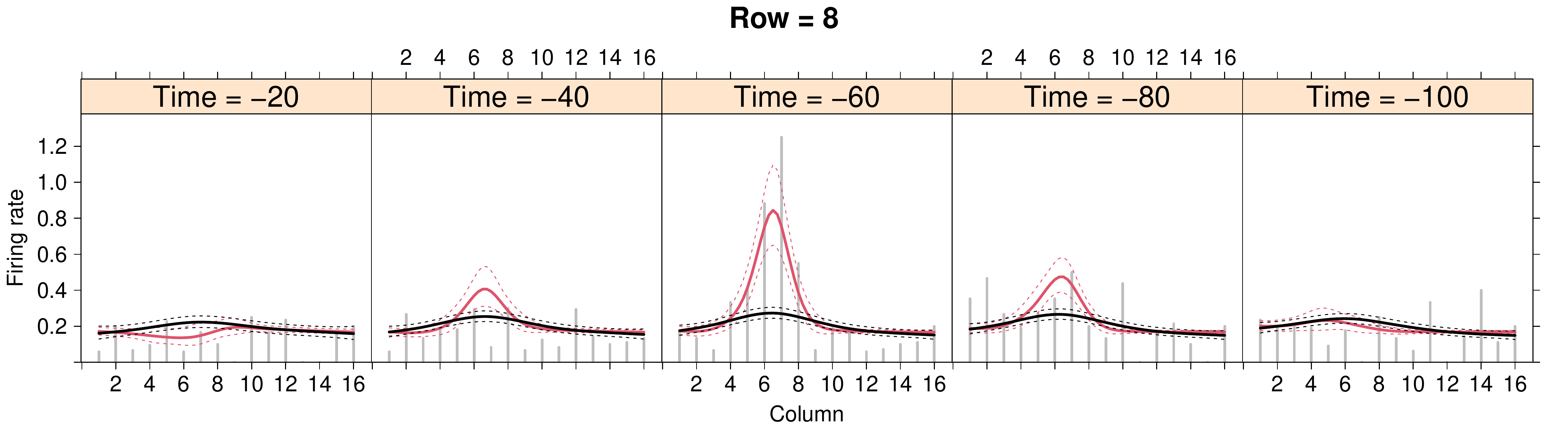}
\caption{\label{MX:FAP2} For the visual receptive field study: observed and smoothed firing rates by row for column $7$ (top figure); and by column for row $8$ (bottom figure). Gray vertical lines: observed. Black line: non-adaptive approach. Red line: adaptive approach. The black and red dotted lines represent 95\% pointwise confidence intervals.}
\end{figure}

We now describe the analysis, and present the results, for the neuron's activity study discussed in Section \ref{motivation}. Recall that the objective is to produce smoothed (de-noised) versions of RFmaps (see Figure \ref{vrf_rawdata}). To that aim, we adopt a Poisson model which expresses the neuronal response as a smooth function of both space (row and column position) and time 
\begin{equation}
\log\left(\mathbb{E}\left[y_{rct} \mid r,c,t\right]\right) = \log\left(n_{rc}\lambda_{rct}\right) = \log\left(n_{rc}\right) + f\left(r,c,t\right),
\label{MX:PoissonModel}
\end{equation}
where $y_{rct}$ denotes the number of spike occurrences attributed to stimulus presentations at row $r$ and column $c$ of the square area ($r,c = 1, \ldots, 16$) for the $t$-th pre-spike time ($t = -20, \ldots, -320$), $n_{rc}$ is the total number of stimulus presentations during the experiment at row $r$ and column $c$, and $\lambda_{rct}$ is the intensity parameter (or firing rate). Model (\ref{MX:PoissonModel}) is estimated with and without assuming locally adaptive smoothing. In both cases, we take the advantage of the array structure of the data and GLAM is used. Besides, we use second-order differences ($q_m = 2$) and marginal cubic B-splines bases of dimension $d_m = 11$ to represent the spatio-temporal surface. For the adaptive approach, we consider the full adaptive penalty and choose $p_{mw} = 6$ ($m,w = 1,2,3$). This gives rise to a total of $648$ ($3\times 6^3$) smoothing parameters (or variance components). Figure \ref{MX:FAP1} depicts (for the same neuron, eye and experimental condition shown in Figure \ref{vrf_rawdata}) the observed as well as the estimated (de-noised) time-series of RFmaps using both approaches. For conciseness, results are not shown for the whole sequence of pre-spike times, but only for those where the RFmaps are structured (the complete results can be found as online material). As can be seen, the adaptive approach shows that the RFmap begins to be structured about $80$ ms pre-spike and peaks approximately at $60$ ms, when a clear central area of high values appears which lasts until $40$ ms. As said, this area represents the visual RF of the neuron. The results indicate that the time between sensory stimulus and neuronal response spans from $40$ to $80$ ms. This is in concordance with the raw data. By contrast, the non-adaptive approach provides too smooth results, and is not able to recover neither the peak nor the temporal structure of the RFmaps (see also Figure \ref{MX:FAP2}). Likewise the previous example, and despite the too smooth results of the non-adaptive approach, the total ED is lower for the adaptive than for the non-adaptive counterpart ($36.4$ and $58.4$, out of $1331$, respectively). In terms of computational effort, in the absence of adaptive smoothing, SOP method takes about $3$ minutes, which increases to $39$ minutes with the adaptive penalty. In contrast to the two-dimensional example of Section \ref{Absenteeism}, here the increase in computing time is substantial. Regarding the cAIC, it suggests that, as expected, a better fit is obtained with the adaptive penalty ($15523.8$ vs $15650.2$ for the standard anisotropic penalty).

\section{Discussion}
%This paper presents a novel approach for multidimensional adaptive penalties in the context of P-spline models. The construction and rationale of the new adaptive penalties are described in full detail for the two-dimensional case, but the extension to three dimensions is also covered. Besides, we discuss several simplifications that can easily be implemented if desired. To the best of our knowledge, the adaptive penalty in three dimensions described in this work represents the first attempt in the literature of P-splines. Model's estimation is based on the equivalence between P-spline models and generalised linear mixed models, and the practical implementation is based on the recently proposed SOP method. However, nothing, in principle, precludes the use of other estimating methods/algorithms provided they can deal with precision/penalty matrices that are linear on the inverse of the variance/smoothing parameters. 
This paper presents a novel approach for anisotropic multidimensional adaptive penalties in the context of P-spline models. The construction and rationale of the new adaptive penalties are described in full detail for the two-dimensional case, but the extension to three dimensions is also covered. Besides, we discuss several simplifications that can easily be implemented if desired. Model's estimation is based on the equivalence between P-spline models and generalised linear mixed models, and the practical implementation is based on the recently proposed SOP method. However, nothing, in principle, precludes the use of other estimating methods/algorithms provided they can deal with precision/penalty matrices that are linear on the inverse of the variance/smoothing parameters. 

Both the simulation study and the applications shown in the paper highlight the gain value of using, when needed, the proposed multidimensional adaptive P-spline models. Also, the simulation study suggests that, when adaptive smoothing is not necessary, the extra flexibility afforded by the adaptive penalties does not translate into an important loss in efficiency. Besides, for most specifications, our adaptive approach outperforms the proposal by \cite{Krivobokova08}, in terms of both MSE and computing time. As an aside, we point out that models based on low-rank radial basis functions and a ridge penalty, such as those used by \cite{Krivobokova08}, do not allow anisotropic smoothing in multiple dimensions, either adaptive or non-adaptive. Of course, the flexibility of multidimensional adaptive P-spline models is at the cost of increasing the computing time. For the two-dimensional case, computing times are kept to a reasonable level, even for rather large B-spline bases and number of smoothing parameters. However, as shown in the applications, the increase is substantial in the three-dimensional case. Even in the non-adaptive case, multidimensional P-splines can be very computationally demanding, especially if large basis dimensions are used. The SOP method \citep[which generalises the proposal presented in][]{MXRA2015} allows alleviating the computational cost, which can otherwise be very large using alternative estimating methods. For instance, using the \texttt{R}-package \texttt{mgcv} \citep{wood2017_book}, estimation of the anisotropic (non-adaptive) P-spline model for the visual receptive field example needs around $21$ minutes. This result seems to indicate that, despite the large computing time required by the three-dimensional adaptive penalty proposed here, estimation would be, if not infeasible, prohibitively computational expensive with other estimating methods than SOP. We note that we used the high-level \texttt{R} language to implement the SOP method for the adaptive approaches described in the paper. It can thus be expected that computing times could be further improved if the method is implemented in a low-level language (e.g. \texttt{C} or \texttt{Fortran}). This would also make computationally more feasible the estimation of more complex models, such as those described in \cite{Rodriguez2012}, where the authors jointly model, for the visual receptive field study, RFmaps under different experimental conditions.

A limitation of the proposal presented here is that, although relaxed, smoothness is still assumed, for both the functions to be estimated and the smoothing parameters. Therefore it is not suitable for functions with abrupt changes or discontinuities \citep[see, e.g.,][for a one-dimensional adaptive approach in these settings]{Liu2010}. The study of appropriate multidimensional penalties (and efficient estimation strategies) for these situations represents an interesting topic for research.   

The \texttt{R}-functions implementing the proposed two- and three- dimensional adaptive P-spline models as well as the \texttt{R}-code used for simulations and the real data examples presented in Section \ref{Application} can be obtained by contacting the first author.
    
\section*{Acknowledgements}
This research was funded by projects MTM2017-82379-R (AEI/FEDER, UE) and PID2019-104901RB-I00 (AEI), by the Basque Government through the BERC 2018-2021 program, by the Spanish Ministry of Science, Innovation, and Universities (BCAM Severo Ochoa accreditation SEV-2017-0718), by Elkartek project 3KIA (KK-2020/00049), by ISCIII (RETICS RD16/0008/0003), Xunta de Galicia (Centro Singular de Investigaci\'on de Galicia, Accreditation 2019-2022, ED431G 2019/02), and the European Regional Development Fund (ERDF).

\appendix
\section{Equivalent Mixed Model for the Adaptive P-spline in Three Dimensions\label{AppA}}
For the three-dimensional case, the mixed model design matrices $\boldsymbol{X}$ and $\boldsymbol{Z}$ (see equations \eqref{MX:2D_X} and \eqref{MX:2D_Z} for the two-dimensional case) are constructed as
\begin{align*}
\boldsymbol{X} & = \boldsymbol{B}\boldsymbol{T}_{0}\\
\boldsymbol{Z} & = \boldsymbol{B}\boldsymbol{T}_{+},
\label{MX:3D_X_Z}
\end{align*}
where $\boldsymbol{B} = \boldsymbol{B}_3\Box\boldsymbol{B}_2\Box\boldsymbol{B}_1$ and
\begin{align*}
\boldsymbol{T}_{0} = {} & [\boldsymbol{U}_{30}\otimes\boldsymbol{U}_{20}\otimes\boldsymbol{U}_{10}],\\
\boldsymbol{T}_{+} = {} & [\boldsymbol{U}_{30}\otimes\boldsymbol{U}_{20}\otimes\boldsymbol{U}_{1+}\mid\boldsymbol{U}_{30}\otimes\boldsymbol{U}_{2+}\otimes\boldsymbol{U}_{10}\mid\boldsymbol{U}_{3+}\otimes\boldsymbol{U}_{20}\otimes\boldsymbol{U}_{10}\mid\\
& \boldsymbol{U}_{30}\otimes\boldsymbol{U}_{2+}\otimes\boldsymbol{U}_{1+}\mid\boldsymbol{U}_{3+}\otimes\boldsymbol{U}_{20}\otimes\boldsymbol{U}_{1+}\mid\boldsymbol{U}_{3+}\otimes\boldsymbol{U}_{2+}\otimes\boldsymbol{U}_{10}\mid\\ 
& \boldsymbol{U}_{3+}\otimes\boldsymbol{U}_{2+}\otimes\boldsymbol{U}_{1+}].
\end{align*}
with $\boldsymbol{U}_{m0}$ and $\boldsymbol{U}_{m+}$ ($m = 1, 2, 3$) defined as in Section \ref{Estimation}. Regarding the precision matrix, it is as follows
\begin{equation*}
\boldsymbol{G}^{-1} = \sum_{u=1}^{p_{11}p_{12}p_{13}}\sigma_{1u}^{-2}\boldsymbol{\mathcal{G}}_{1u} + \sum_{s=1}^{p_{21}p_{22}p_{23}}\sigma_{2s}^{-2}\boldsymbol{\mathcal{G}}_{2s} + \sum_{v=1}^{p_{31}p_{32}p_{33}}\sigma_{3v}^{-2}\boldsymbol{\mathcal{G}}_{3v},
\end{equation*}
where $\sigma_{1u}^{2} = \phi/\xi_{1u}$, $\sigma_{2s}^2 = \phi/\xi_{2s}$, $\sigma_{3v}^2 = \phi/\xi_{3v}$ and 
\begin{align*}
\boldsymbol{\mathcal{G}}_{1u} & = \boldsymbol{T}_{+}^{\top}\left(\mathbf{I}_{d_3}\otimes\mathbf{I}_{d_2}\otimes\boldsymbol{D}_{q_1}\right)^{\top}\mbox{diag}\left(\boldsymbol{\psi}_{1,u}\right)\left(\mathbf{I}_{d_3}\otimes\mathbf{I}_{d_2}\otimes\boldsymbol{D}_{q_1}\right)\boldsymbol{T}_{+},\\
\boldsymbol{\mathcal{G}}_{2s} & = \boldsymbol{T}_{+}^{\top}\left(\mathbf{I}_{d_3}\otimes\boldsymbol{D}_{q_2}\otimes\mathbf{I}_{d_1}\right)^{\top}\mbox{diag}\left(\boldsymbol{\psi}_{2,s}\right)\left(\mathbf{I}_{d_3}\otimes\boldsymbol{D}_{q_2}\otimes\mathbf{I}_{d_1}\right)\boldsymbol{T}_{+},\\
\boldsymbol{\mathcal{G}}_{3v} & = \boldsymbol{T}_{+}^{\top}\left(\boldsymbol{D}_{q_3}\otimes\mathbf{I}_{d_2}\otimes\mathbf{I}_{d_1}\right)^{\top}\mbox{diag}\left(\boldsymbol{\psi}_{3,v}\right)\left(\boldsymbol{D}_{q_3}\otimes\mathbf{I}_{d_2}\otimes\mathbf{I}_{d_1}\right)\boldsymbol{T}_{+}.
\end{align*}

\bibliographystyle{hapalike}
\bibliography{m_SOP}
\end{document}